%% file: paper_R2.tex
\documentclass[micromachines,article,submit,moreauthors,pdftex,12pt,a4paper]{mdpi}

\usepackage{graphicx}
\usepackage{amsmath,amssymb}
\usepackage{hyperref}

\newcommand{\newtext}[1]{#1}

\include{format}

\lastpage{x}
\doinum{10.3390/------}
\pubvolume{xx}
\pubyear{2010}
\history{Received: xx / Accepted: xx / Published: xx}

\pdfoutput=1

\Title{Mori-Tanaka Based Estimates of Effective Thermal Conductivity of
Various Engineering Materials}

\Author{Jan Str\'{a}nsk\'{y}$^{1}$, Jan Vorel$^{1}$, Jan Zeman$^1$ and
Michal \v{S}ejnoha$^{1,2,\star}$}

\address{%
$^{1}$ Department of Mechanics, Faculty of Civil Engineering, Czech Technical
  University in Prague, Prague, Czech Republic\\
$^{2}$ Centre for Integrated Design of Advanced
  Structures, Faculty of Civil Engineering, Czech Technical University in
  Prague, Prague, Czech Republic
}

\corres{E-Mail: sejnom@fsv.cvut.cz, Tel.: +420-2-2435-4494 and
Fax:+420-2-2435-4494}

\abstract{The purpose of this paper is to present a simple micromechanics-based
model to estimate the effective thermal conductivity of macroscopically
isotropic materials of matrix-inclusion type. The methodology is based on the
well-established Mori-Tanaka method for composite media reinforced with
ellipsoidal inclusions, extended to account for imperfect thermal contact at the
matrix-inclusion interface, random orientation of particles and particle size
distribution. Using simple ensemble averaging arguments, we show that the
Mori-Tanaka relations are still applicable for these complex systems, provided
that the inclusion conductivity is appropriately modified.  Such conclusion is
supported by the verification of the model against a detailed finite-element
study as well as its validation against experimental data for a wide range of
engineering material systems.}


\begin{document}

\section{Introduction}\label{sec:Introduction}
There has been a clear trend over the last decade to exploit ever greater detail
of the material structure towards better predictions of its response from
simulations. Hierarchical modeling strategy, regardless whether coupled or
uncoupled but mostly of the bottom-up type, has served to provide estimates of
the macroscopic response. In this process, geometric details decisive for a
given scale are first quantified employing various statistical
descriptors~\cite{Torquato:2002:RHM}, but eventually smeared via homogenization
to render larger scale property. Greater precision is expected when introducing
the results of microstructure evaluation into the homogenization step. However,
the actual gain when compared to the cost of this analysis is still in question.
Obviously, description of evolving microstructures or rigorous representation of
deformation mechanisms would require to account for almost every detail of the
microstructure on a given scale. But how deep do we have to go if only the
\newtext{effective macroscopic} response (i.e. linear macroscopic properties) is
of the primary interest? Such a goal is \newtext{addressed} in this contribution.

\begin{figure}[ht]
\begin{center}
\begin{tabular}{c@{\hspace{10mm}}c@{\hspace{10mm}}c}
\includegraphics*[width=50mm,keepaspectratio]{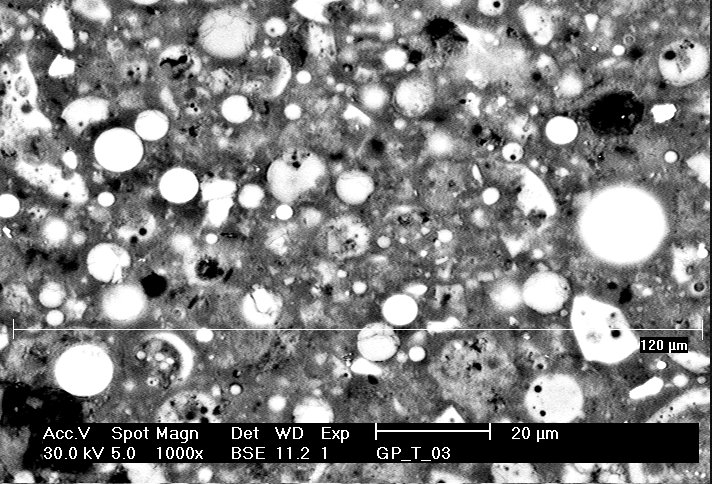}&
\includegraphics*[width=50mm,height=37.5mm]{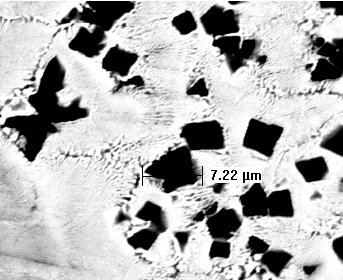}&
\includegraphics*[width=50mm,keepaspectratio]{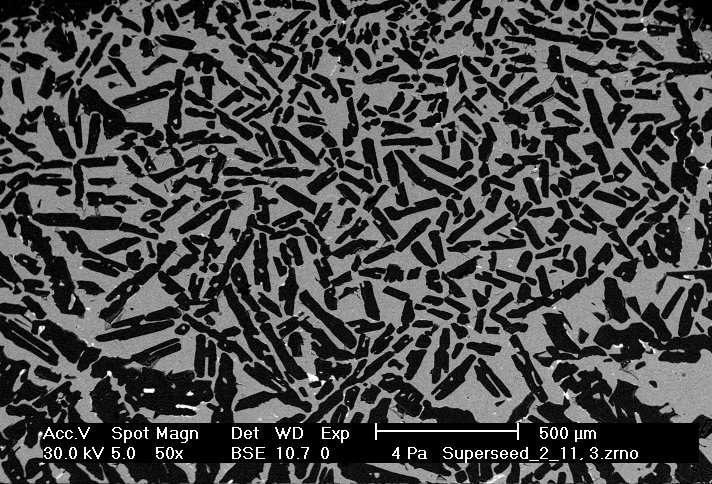}\\
(a)&(b)&(c)
\end{tabular}
\end{center}
\caption{Examples of micro-graphs of real engineering materials taken
  in back scattered electrons: a) Alkali-activated fly
  ash, b) Alumino-silicate ceramics with Fe and
  silicium particles (dark phase), c)~Superspeed - alloying ingredient
  into crude iron for cast iron working with silicon particles (dark
  phase). \newtext{Reproduced with permission of} L. Kopeck\'{y} (CTU in
  Prague).}
\label{fig:eng-mat}
\end{figure}

Here, the modeling effort concentrates on the evaluation of effective
thermal conductivities of various engineering materials with a
significant degree of heterogeneity whereas focusing on imperfect
thermal contact along constituents interfaces. We shall argue,
shielded by available experimental data, that reasonably accurate
predictions of macroscopic response can be obtained with very limited
information about actual microstructure such as volume fractions and
local properties of material phases. Consequently, we lump the entire
analysis on the assumption of representing true material structures by
statistically isotropic distribution of spheres.~\figref{fig:eng-mat}
shows micro-images of selected material representatives which seem to
admit this classification. Note that whatever material phase embedded
into the matrix (the basic material) is henceforth termed the
heterogeneity in real material systems while it is termed
inclusion in approximations adopted for calculations.

\begin{figure}[hbtp]
\begin{center}
\begin{tabular}{c@{\hspace{10mm}}c}
\includegraphics*[height=55mm]{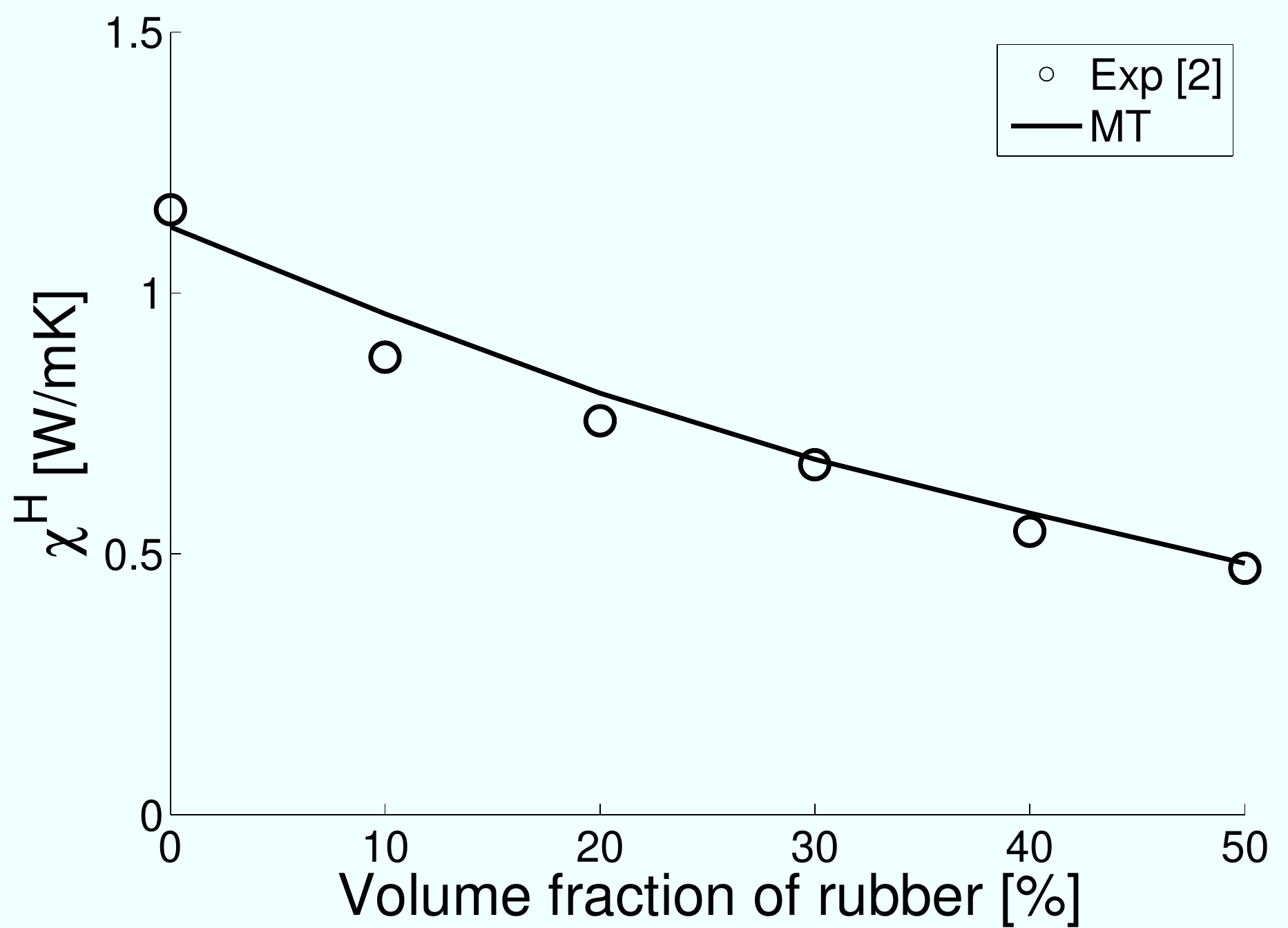}&
\includegraphics*[height=55mm]{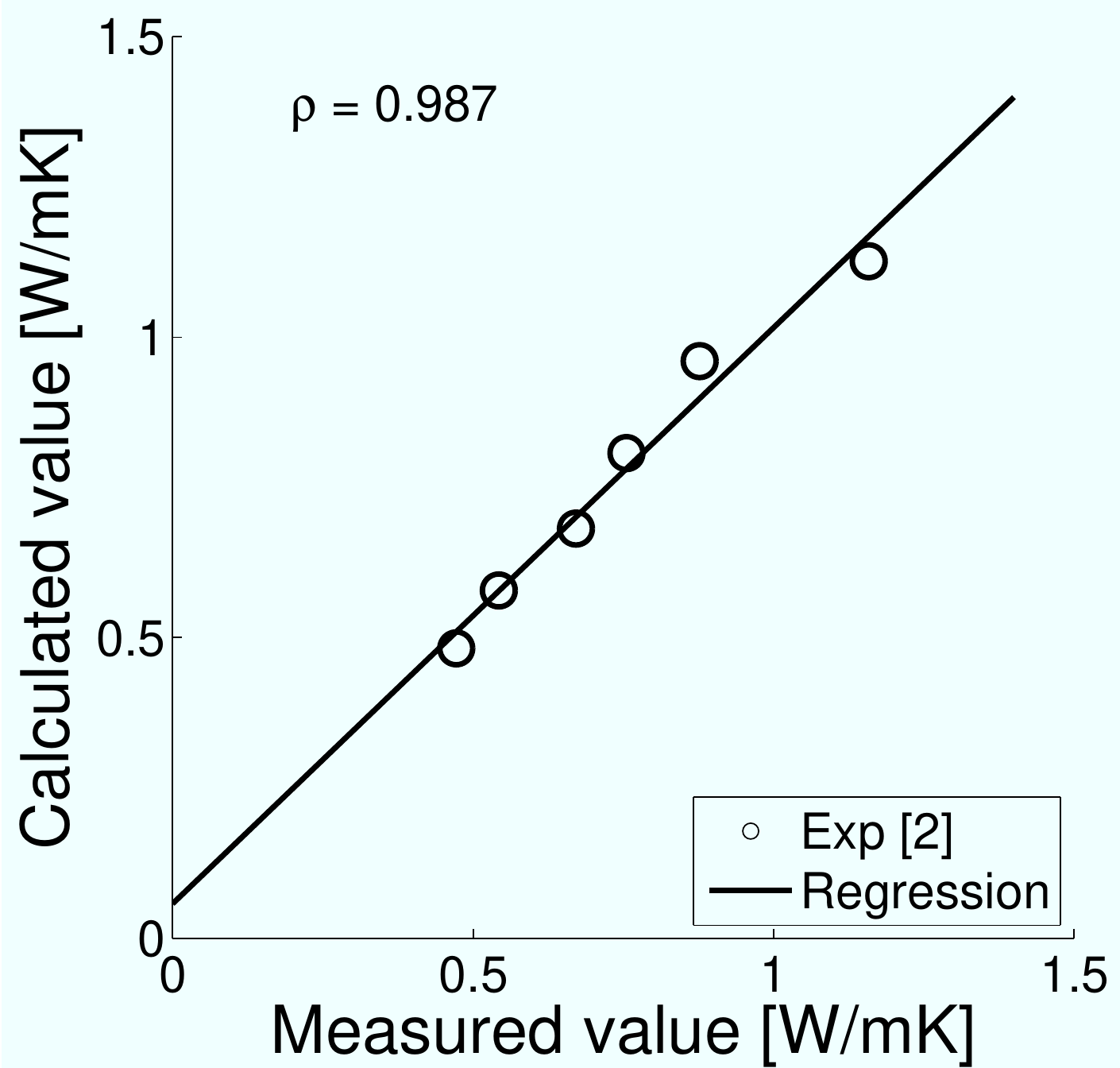}\\
(a)&(b)
\end{tabular}
\end{center}
\caption{a) Evolution of effective thermal conductivity \newtext{$\KH$} as a
function of volume fraction of rubber in solid phase, b) correlation of
  measured and calculated values; \newtext{$\rho$ is the correlation
  coefficient}.}
\label{R_grafy}
\end{figure}

Strong motivation for this seemingly swingeing simplification is
supported by experimental measurements presented in~\cite{rubber} for
cement matrix based mixture of rubber particles and air
voids. Comparison between experimental data and predictions provided
by the Mori-Tanaka averaging scheme under the premise of random
distribution of spherical inclusions, the method in this particular
format promoted herein, appears in~\figref{R_grafy}. The match is
almost remarkable.

Going back to~\figref{fig:eng-mat} one may object that while
admissible for~\figref{fig:eng-mat}(a) the spherical approximation of
particle phase in~\figref{fig:eng-mat}(c) will yield rather erroneous
predictions. Note, however, that this attempt is not hopeless
providing the microstructure can still be considered as
macroscopically isotropic, ensured for statistically isotropic
distribution of heterogeneities having isotropic material symmetry.
In that case, it can be shown that the previously mentioned
Mori-Tanaka method written out for spherical inclusions is adequate
providing the material properties of the inclusions are suitably
modified. Although this step requires information beyond that of
volume fractions of phases, 
the benefit of gathering additional data will become particularly
appreciable once turning our attention to material systems with
imperfect interfaces, the principal objective of this study.

The problem of quantifying the influence of imperfect thermal contact on the
overall thermal conductivity has been under intense study in the past. Hasselman
and Johnson~\cite{Hasselman} provided estimates for dilute concentration of
mono-disperse spherical and cylindrical heterogeneities. Successful application
of this simple model to Al/SiC porous composites is presented in~\cite{AlSi}.
The Hasselman-Johnson results were then extended by Benveniste and
Miloh~\cite{Benve2} to spheroidal particle shapes with imperfect interfaces and
subsequently applied in the framework of the Mori-Tanaka method~\cite{Benve}.
These early developments were later generalized by Nogales and B\"{o}hm, who
proposed in~\cite{Bohm} a simple method for dealing with polydisperse systems of
spherical particles. In addition, rigorous third-order bounds for effective
conductivity of \newtext{macroscopically isotropic distribution of particles}
with imperfect interfaces were derived by Torquato and Rintoul~\cite{Tor}.
Alternatively, as demonstrated by Hashin~\cite{Hashin:2001:TII}, the material
systems with imperfect interfaces can be accurately approximated by the coated
inclusion model due to Dunn and Taya~\cite{Dun:1993:ETC}, which also accounts
for different orientation of the inclusions. If limiting attention to spherical
inclusions the results presented in~\cite{Bohm} can be obtained in \newtext{a}
very elegant way by simple extension of one-dimensional analysis. This is
demonstrated in Appendix~\ref{sec:appb}.

\begin{figure}[ht]
\begin{center}
\includegraphics[height=80mm]{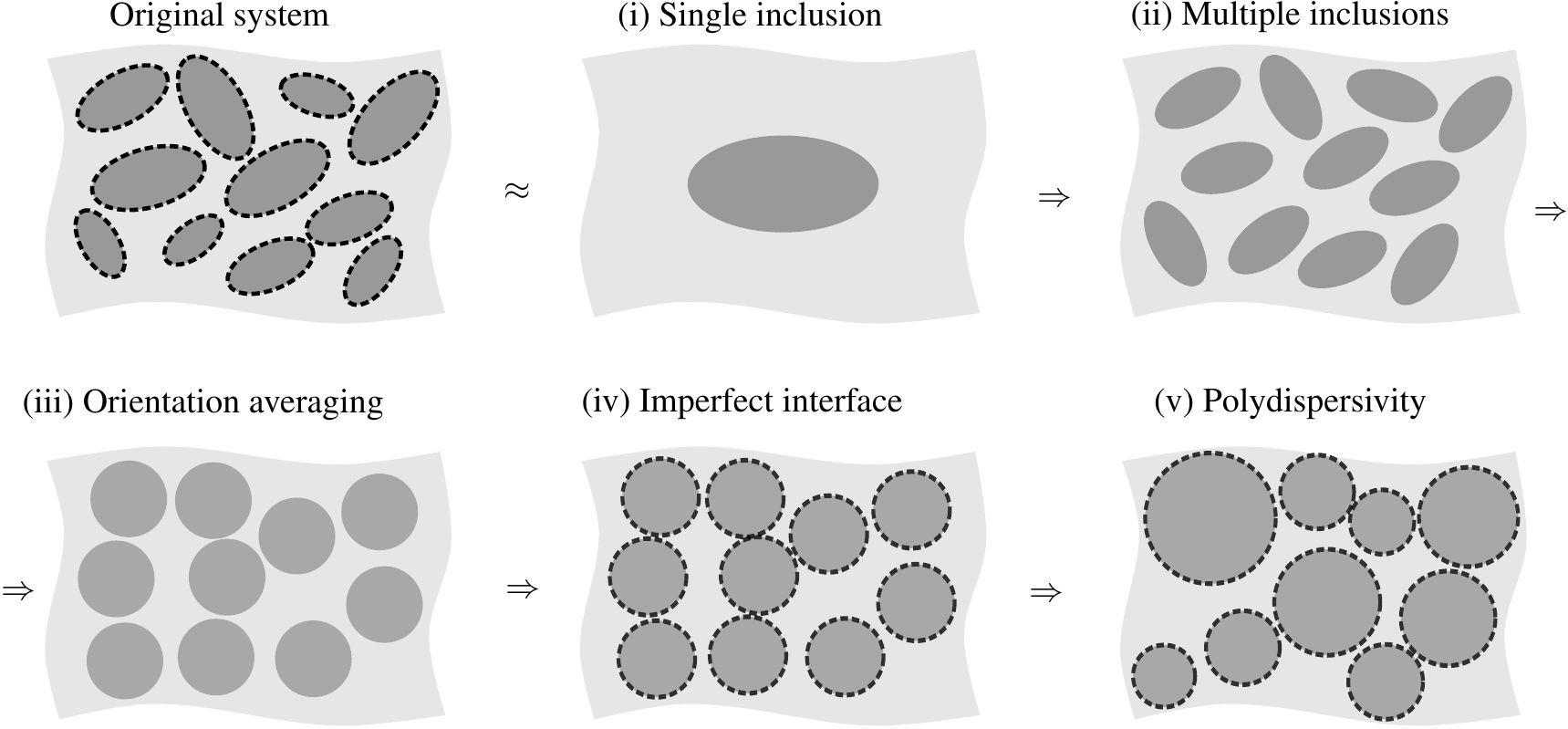}
\end{center}
\caption{Mori-Tanaka based scheme: Strategy of derivation}
\label{fig:derivation_strategy}
\end{figure}

To exploit this result in practical applications of the Mori-Tanaka
method to a heat conduction problem, prediction of effective thermal
conductivity in particular, we adopt the analysis scheme graphically
presented in~\figref{fig:derivation_strategy}. We start from the
assumption of multidisperse system of randomly oriented spheroidal
inclusions with possibly imperfect thermal contact (non-zero
temperature jump along the interface). To arrive at the desired
approximation of multidisperse system of spherical inclusions with
perfect interfaces (temperature continuity along the interface) we
proceed in five consecutive steps.

These steps are mathematically described
in~\secref{sec:MTthoery}.~\secref{sec:MTexamples} is devoted to both
validation and verification of the proposed scheme against available
experimental data and finite element simulations performed for several
representatives of statistically isotropic random microstructures. The
crucial results and principal recommendations are finally
summarized in~\secref{sec:concl}.

\section{Theoretical background of the Mori-Tanaka method}\label{sec:MTthoery}
In this section attention is accorded to essential theoretical details of the
Mori-Tanaka method in view of the five steps
in~\figref{fig:derivation_strategy}. In the first step, we consider a single
inclusion with perfect interface subject to far-field loading. This step is
theoretically elaborated in~\secref{sec:SingleInhomogeneity}.  Solution of this
problem is then employed in~\secref{sec:MultipleInhomogeneities} to estimate the
overall conductivity of a composite consisting of multiple ellipsoidal
inclusions bonded to a matrix phase. The third step addressed
in~\secref{sec:OrientationAveraging} is reserved for systems with randomly
oriented inclusions with a uniform distribution over the hemisphere. Here, a
simple orientation averaging argument is shown to demonstrate that the effective
conductivity of the system coincides with the conductivity of a system
reinforced by spherical inclusions, thermal conductivity \newtext{of which} is
appropriately modified. \newtext{Following~\cite{Bohm}}, an analogous argument is employed in
the next step outlined in~\secref{sec:ImperfectInterface} to account for imperfect
thermal contact along the matrix-inclusion interface. It is shown that in this
case the modified conductivity becomes size-dependent. This eventually allows us
to extend the scheme to polydisperse systems
in~\secref{sec:PolydisperseSystems}.

\subsection{Single inclusion with perfect interface}\label{sec:SingleInhomogeneity}
Let us consider an ellipsoidal inclusion $\Oi$, with semi-axes \newtext{$a_1
\leq a_2 \leq a_3$} embedded in the matrix domain $\Om$. We attach to the
inclusion a Cartesian coordinate system with the origin at the
inclusion center and axes aligned with the semi-axes. The distribution
of local fields follows from the problem
\begin{eqnarray}\label{eq:fourier}
\mQ( \x )
=
-
\mK( \x )
\mH( \x ),
& 
\vek{\nabla}\trn
\mQ( \x )
=
0
&
\mbox{for } \x \in \mathbb R^3,
\end{eqnarray}
where $\mQ \in \mathbb R^3$ denotes the heat flux, $\mH \in \mathbb
R^3$ denotes gradient of the temperature $\tmp$~(i.e. $\mH(\x) =
\vek{\nabla}\tmp(\x)$) and $\mK$ designates the $3 \times 3$ symmetric
positive-definite matrix of thermal conductivity given by
\begin{equation}
\mK( \x )
= 
\left\{
\begin{array}{cl}
\mKi & \mbox{for } \x \in \Oi, \\
\mKm & \mbox{otherwise.}
\end{array}
\right.
\end{equation}
Eq.~\eqref{eq:fourier} is completed by the far-field boundary
conditions, cf.~\cite[Eq.~(14)]{Benve2},
\begin{eqnarray}\label{eq:far_field_bcs}
\tmp( \x )
=
\MH \trn \x
&&
\mbox{for } \| \x \| \rightarrow \infty,
\end{eqnarray}
with $\MH \in \mathbb R^3$ denoting the overall~(macroscopic)
temperature gradient. Due to linearity of the problem, we can
introduce the temperature gradient concentration factor $\mCH \in
\mathbb R^{3 \times 3}$ in the form:
\begin{eqnarray}
\mH( \x )
= 
\mCH( \x )
\MH
&&
\mbox{for } \x \in \mathbb{R}^3.
\label{eq:CF}
\end{eqnarray}
As shown first by Hatta and Taya~\cite{EIM}, the concentration factor
is constant inside the inclusion and admits the expression
\begin{eqnarray}\label{eq:conc_fact}
\mCH^{-1}( \x )
=
\left( \mCHi \right)^{-1}
=
\I
-
\Et
\left( \mKm \right)^{-1}
\left( \mKm - \mKi \right)
&&
\mbox{for } \x \in \Oi,
\end{eqnarray}
where $\I$ denotes the unit matrix and $\Et \in \mathbb R^{3\times 3}$ is the
Eshelby-like tensor which depends only \newtext{on} the matrix conductivity
$\mKm$ and the ratios of semi-axes lengths $\newtext{\ar_2} = a_2 : a_1$ and
$\newtext{\ar_3} = a_3 : a_1$, see also Appendix~\ref{sec:appa} for additional
details. For the spherical inclusion with isotropic conductivity $\mKi = \Ki\I$
embedded in \newtext{the} isotropic matrix with $\mKm = \Km\I$,
Eq.~\eqref{eq:conc_fact} simplifies \newtext{as}
\begin{eqnarray}
\mCHi
=
\CHi\sph
\I
& \mbox{with} &
\CHi\sph 
= 
\frac{%
3 \Km
}{%
2 \Km + \Ki
}.
\end{eqnarray}

\subsection{Multiple inclusions with perfect interface}\label{sec:MultipleInhomogeneities}
In the next step, we adopt the results of the previous section to
estimate the overall behavior of a composite material consisting of
distinct phases $r=0,1,\ldots, N$. The value $r=0$ is reserved for the
matrix phase $\Omega^{\mtrx}$ and the $r$-th phase ($r>0$) corresponds
to the ellipsoidal inclusion $\Omega\phs{r}$, characterized by its
semi-axes $a\phs{r}_1, a\phs{r}_2$ and $a\phs{r}_3$, volume fraction
$\vfrac\phs{r}$ and conductivity $\mK\phs{r}$. Following
Benveniste's reformulation~\cite{Benvensite:1987:MTM} of the original
Mori-Tanaka scheme~\cite{Mori:1973:ASM}, the interaction among phases
is approximated by subjecting each inclusion separately to the mean
temperature gradient in the matrix phase $\MHm$ in
Eq.~\eqref{eq:far_field_bcs}. As a result, the temperature gradient
inside the $r$-th phase remains constant and reads
\begin{eqnarray}\label{eq:Hr}
\MH\phs{r}
=
\mCHr{r}
\MHm
&\mbox{for}&
r = 0, 1, \ldots, N.
\end{eqnarray}
Here, 
\begin{equation}
\MH\phs{r}
=
\frac{1}{\meas{\O\phs{r}}}
\int_{\O\phs{r}}
\mH( \x )
\de \x,
\end{equation}
and $\mCHr{r}$ is \newtext{the} $3\times 3$ partial temperature gradient
concentration factor of the $r$-th phase, given by
\begin{equation}
\mCHr{r}
=
\left\{
\begin{array}{cl}
\I 
& \mbox{for } r=0, 
\\
\vek{R}\phs{r}
\mCHr{r}\loc
(\vek{R}\phs{r})^{\trn}
& \mbox{for } r = 1,2, \ldots, N,
\end{array}
\right.
\end{equation}
where the $3\times 3$ rotation matrix $\vek{R}\phs{r}$ accounts for
the difference in the global and local coordinate systems, see
\secref{sec:OrientationAveraging} for additional details, and
$\mCHr{r}\loc$ equals to $\mCHi$ in Eq.~\eqref{eq:conc_fact} with $\mKi
= \mK\phs{r}$ and $\Et$ determined from values $a\phs{r}_1,
a\phs{r}_2$ and $a\phs{r}_3$.

The volume consistency of the overall temperature gradient $\MH$ and
local averages $\MH\phs{r}$ requires
\begin{equation}
\MH
=
\sum_{r=0}^N
\vfrac\phs{r}
\MH\phs{r}
=
\left( 
\sum_{r=0}^N
\vfrac\phs{r}
\mCHr{r}
\right)
\MHm.
\end{equation}
Inverting the above equation gives the average temperature gradient in the matrix as
\begin{equation}
\MHm = 
\left( 
\sum_{r=0}^N
\vfrac\phs{r}
\mCHr{r}
\right)^{-1}
\MH
\,=\, \mCH^{\mtrx}\MH,
\end{equation}
which, when substituted into Eq.~\eqref{eq:Hr}, yields the explicit
expression for the phase temperature gradients in the form
\begin{eqnarray}
\MH\phs{r}
=
\mCHr{r}
\left( 
\sum_{r=0}^N
\vfrac\phs{r}
\mCHr{r}
\right)^{-1}
\MH
\,=\, \mCH\phs{r}\MH,
\end{eqnarray}
where $\mCH^{\mtrx}, \mCH\phs{r}$ are the matrix and inclusion
temperature gradient concentration factors, respectively.

As each phase is assumed to be homogeneous, the average heat flux in
the $r$-th phase
\begin{eqnarray}
\MQ\phs{r}
=
\frac{1}{\meas{\O\phs{r}}}
\int_{\O\phs{r}}
\mQ( \x )
\de \x
&&
\mbox{for } 
r = 0, 1, \ldots, N,
\end{eqnarray}
equals to
\begin{eqnarray}
\MQ\phs{r}
=
- \mK\phs{r}
\MH\phs{r}
&&
\mbox{for }
r = 0,1, \ldots, N.
\end{eqnarray}
This allows us to express the macroscopic heat flux in the form
\begin{equation}
\MQ
=
\sum_{r=0}^N
\vfrac\phs{r}
\MQ\phs{r}
=
-
\left(
\sum_{r=0}^N
\vfrac\phs{r}
\mK\phs{r}
\mCHr{r}
\right)
\left( 
\sum_{r=0}^N
\vfrac\phs{r}
\mCHr{r}
\right)^{-1}
\MH,
\end{equation}
from which we obtain the effective conductivity $\MQ = - \MK\MH$ in
its final form
\begin{equation}\label{eq:eff_cond_MT}
\MK
=
\left(
\vfrac^{\mtrx}
\mKm
+
\sum_{r=1}^N
\vfrac\phs{r}
\mK\phs{r}
\mCHr{r}
\right)
\left( 
\vfrac^{\mtrx} 
\I
+
\sum_{r=1}^N
\vfrac\phs{r}
\mCHr{r}
\right)^{-1}.
\end{equation}
Finally, assuming that the composite consists of isotropic matrix
with conductivity $\Km$ and spherical isotropic inclusions with
conductivities $\K\phs{r}$, Eq.~\eqref{eq:eff_cond_MT} becomes $\MK =
\KH \I$, with
\begin{eqnarray}\label{eq:eff_conductivity_shperical}
\KH
=
\frac{%
\vfrac^{\mtrx}
\Km
+
\displaystyle{\sum_{r=1}^N}
\vfrac\phs{r}
\K\phs{r}
\pCH{r}\sph
}{
\vfrac^{\mtrx} 
+
\displaystyle{\sum_{r=1}^N}
\vfrac\phs{r}
\pCH{r}\sph
}
&\mbox{and}&
\pCH{r}\sph
= 
\frac{%
3 \Km
}{%
2 \Km + \K\phs{s}}
.
\end{eqnarray}

\subsection{Orientation averaging}\label{sec:OrientationAveraging}
We are now in a position to provide estimates of the effective thermal
conductivity for composites with $M$~(with $M \ll N$)
inclusion classes indexed by $s = 1, 2, \ldots, M$. Each class is
characterized by a single Eshelby-like matrix $\Et$ in
Eq.~\eqref{eq:conc_fact} and represents the reference ellipsoidal
inclusion randomly oriented over the unit hemisphere with an
independent uniform distribution of orientation angles.

To this goal, consider a quantity $\vek{X}\loc \in \mathbb{R}^{3
  \times 3}$, expressed in a local coordinate system aligned with a
certain reference inclusion. Its form in the global coordinate system
follows from
\begin{equation}
\vek{X}(\newtext{\varphi}, \newtext{\phi}, \newtext{\psi}) 
= 
\vek{R}( \newtext{\varphi}, \newtext{\phi}, \newtext{\psi} )
\vek{X}\loc
\vek{R}\trn( \newtext{\varphi}, \newtext{\phi}, \newtext{\psi} ),
\end{equation}
where $\newtext{\varphi}, \newtext{\phi}$ and $\newtext{\psi}$ denote the Euler
angles\footnote{Note that so-called "$x_2$ convention" is used, in
  which a conversion into a new coordinates system follows three
  consecutive steps. First, the rotation of angle $\newtext{\varphi}$ around the
  original $X_3$ axis is done. Then, the rotation of angle $\newtext{\phi}$
  around the new $x_2$ axis is followed by the rotation of angle
  $\newtext{\psi}$ around the new $x_3$ axis to finish the conversion.} and
the transformation matrix $\vek{R}$ is provided by
\begin{equation}
\vek{R}( \newtext{\varphi}, \newtext{\phi}, \newtext{\psi} )
=
\left[
\begin{array}{rrr}
  \cos\newtext{\psi} & 
 -\sin\newtext{\psi} & 
  0 \\
  \sin\newtext{\psi} & 
  \cos\newtext{\psi} &
  0 \\
  0 & 0 & 1
\end{array}
\right]
\left[
  \begin{array}{rrr}
  \cos\newtext{\phi} & 0 & -\sin\newtext{\phi} \\
  0 & 1 & 0 \\
  \sin\newtext{\phi} & 0 & \cos\newtext{\phi} 
  \end{array}
\right]
\left[
  \begin{array}{rrr}
  \cos\newtext{\varphi} & -\sin\newtext{\varphi} & 0 \\
  \sin\newtext{\varphi} &  \cos\newtext{\varphi} & 0 \\
  0 & 0 & 1
  \end{array}
\right].
\end{equation}
The orientation average of $\vek{X}$ is denoted by double angular
brackets:
\begin{equation}
\oavg{\vek{X}}
=
\frac{1}{8\pi^2}
\int_0^{2\pi} 
  \int_0^\pi 
    \int_0^{2\pi}
    \vek{X}(\newtext{\varphi}, \newtext{\phi}, \newtext{\psi})
    \sin \newtext{\phi}
    \de\newtext{\varphi} 
  \de\newtext{\phi} 
\de\newtext{\psi}.
\end{equation}
Straightforward calculation, presented e.g. in~\cite[Appendix
  A.2.3]{Stransky:2009:MMTC}, reveals that the orientation averaging
of an arbitrary $\vek{X}\loc \in \mathbb R^{3\times 3}$ yields
\begin{eqnarray}
\oavg{\vek{X}}
=
\oavg{X} \I
&\mbox{with}&
\oavg{X} 
= 
\frac{1}{3} 
\sum_{i=1}^3 
\left( X\loc\right)_{ii}.
\end{eqnarray}

Repeating the steps of the previous section with partial temperature
gradient concentration factors replaced with their orientation
averages, we obtain, after some manipulations presented e.g.
in~\cite[Section B]{Benveniste:1990:ETC}, the scalar homogenized
conductivity in the form
\begin{equation}\label{eq:eff_conductivity_averaged}
\KH
=
\frac{%
\vfrac^{\mtrx}
\Km
+
\displaystyle{\sum_{s=1}^M}
\vfrac\phs{s}
\oavg{\K\phs{s}\pCH{s}}
}{%
\vfrac^{\mtrx}
+
\displaystyle{\sum_{s=1}^M}
\vfrac\phs{s}
\oavg{\pCH{s}}
}.
\end{equation}
Therefore, it follows from the comparison of
Eq.~\eqref{eq:eff_conductivity_averaged} with
Eq.~\eqref{eq:eff_conductivity_shperical} that the system of randomly
oriented inclusions embedded in an isotropic matrix is
indistinguishable, from the point of view of homogenized conductivity,
from the system of spherical inclusions with an \emph{apparent}
conductivity
\begin{equation}\label{eq:apparent_cond_def}
\widetilde{\K}\phs{s}
=
\frac{\oavg{\K\phs{s}\pCH{s}}}{\oavg{\pCH{s}}},
\end{equation}
which yields
\begin{eqnarray}\label{eq:eff_isotropic_conductivity}
\KH
=
\frac{%
\vfrac^{\mtrx}
\Km
+
\displaystyle{\sum_{s=1}^M}
\vfrac\phs{s}
\widetilde{\K}\phs{s}
\widetilde{T}\phs{s}\sph
}{%
\vfrac^{\mtrx}
+
\displaystyle{\sum_{s=1}^M}
\vfrac\phs{s}
\widetilde{T}\phs{s}\sph
}
&\mbox{with}&
\widetilde{T}\phs{s}\sph
= 
\frac{%
3 \Km
}{%
2 \Km + \widetilde{\K}\phs{s}}.
\end{eqnarray}

\subsection{Imperfect interface}\label{sec:ImperfectInterface}
The presence of imperfect thermal contact at the matrix-inclusion
interface $\partial\Oi$ results in temperature jump $\jump{\tmp}$,
\newtext{the magnitude of which} is provided by Newton's law,
e.g.~\cite[Section 1.3]{MIT}:
\begin{eqnarray}
\n\trn( \x )
\mQ( \x )
=
k( \x )
\jump{\tmp( \x )}
&&
\mbox{for } \x \in \partial \Oi,\label{eq:temp_jump}
\end{eqnarray}
where $k$ denotes the interfacial conductance (with $k \rightarrow
\infty$ corresponding to perfect interface and $k = 0$ to ideal
insulation) and $\n$ denotes the normal vector at the interface
oriented outside the inclusion. This relation, together with
Eqs.~\eqref{eq:fourier}--\eqref{eq:far_field_bcs}, defines the single
inclusion problem accounting for the presence of imperfect
interface. Its solution is, however, substantially more involved as
the temperature gradient inside an ellipsoidal inclusion becomes
position-dependent; the concentration factor is then available only in
the form of complicated infinite series expansion for spheroidal
inclusions~\cite{Benve2} or ellipsoidal coated
inclusions~\cite{Dun:1993:ETC}. Nevertheless, when restricting the
attention to spherical inclusion of radius $a$, it can be shown that
the temperature gradient within inclusion recovers the constant value
and the concentration factor becomes~\cite{Benve2}
\begin{eqnarray}\label{eq:single_particle_imperfect}
\widehat{\CH}\sph( \Ki, k, \rd )
=
\frac{%
3 \Km
}{%
2 \Km + \widehat{\K}^\incl( \Ki, k, \rd )
}
&\mbox{where}&
\widehat{\K}^\incl( \Ki, k, \rd )
=
\Ki
\frac{ak}{ak + \Ki},
\end{eqnarray}
see also Appendix~\ref{sec:appb} for simple derivation of this
result. Hence, analogously to the previous section, the interfacial
effect can be modeled by replacing the ``true'' conductivity $\Ki$ by
the \emph{size-dependent} apparent value $\widehat{\K}^\incl$ provided
by Eq.~\eqref{eq:single_particle_imperfect}$_2$. Assuming in addition
that each class of inclusions is characterized by identical semi-axes
lengths $a_1\phs{s}, a_2\phs{s}$ and $a_3\phs{s}$ and interfacial
conductance $k\phs{s}$, we propose to extend the
relation~\eqref{eq:eff_isotropic_conductivity} into the form,
cf.~\cite[Section 2]{Bohm}
\begin{eqnarray}\label{eq:general_imperfect_system}
\KH
=
\frac{%
\vfrac^{\mtrx}
\Km
+
\displaystyle{\sum_{s=1}^M}
\vfrac\phs{s}
\widehat{\K}\phs{s}
\widehat{T}\phs{s}\sph
}{%
\vfrac^{\mtrx}
+
\displaystyle{\sum_{s=1}^M}
\vfrac\phs{s}
\widehat{T}\phs{s}\sph
},
&
\displaystyle
\iCH{s}\sph
= 
\frac{%
3 \Km
}{%
2 \Km + \widehat{\chi}\phs{s}
}
,
&
\widehat{\K}\phs{s}
=
\widetilde{\K}\phs{s}
\frac{a\phs{s} k\phs{s}}{a\phs{s} k\phs{s} + \widetilde{\K}\phs{s}},
\end{eqnarray}
with $\rd\phs{s} = \sqrt[3]{a\phs{s}_1 a\phs{s}_2 a\phs{s}_3}$ and the
apparent conductivity $\widetilde{\K}\phs{s}$ given by
Eq.~\eqref{eq:apparent_cond_def}.

\subsection{Polydisperse systems}\label{sec:PolydisperseSystems}
Even though Eq.~\eqref{eq:general_imperfect_system} is applicable to
very general material systems, in practice we typically assume single
inclusion family, with the particle size distribution characterized by
a probability density function $p(a)$ satisfying
\begin{eqnarray}
p(a) \geq 0 \mbox{ for } -\infty < a < \infty,
&&
\int_{-\infty}^{\infty}
p(a) \d a = 1. 
\end{eqnarray}
In this context, the effective conductivity finally becomes
\begin{equation}\label{eq:polydisp}
\KH
=
\frac{%
\vfrac^{\mtrx}
\Km
+
\vfrac\phs{1}
\savg{
\widehat{\K}\phs{1}
\widehat{T}\phs{1}\sph
}
}{%
\vfrac^{\mtrx}
+
\vfrac\phs{1}
\savg{\widehat{T}\phs{1}\sph}
},
\end{equation}
where, for \newtext{an} arbitrary function $g(a)$, $\savg{g}$ denotes its
expected value given by
\begin{equation}\label{eq:integ}
\savg{g} 
= 
\int_{-\infty}^{\infty}
g(a) p(a) \d a.
\end{equation}
Following e.g.~\cite{Molina:2005:LMI,Bohm}, the log-normal
distribution with the probability density function
\begin{eqnarray}\label{eq:log-normal}
p( \rd )
=
\frac{1}{\sqrt{2\pi} a \sigma}
\exp\left( 
-\left[
  \frac{\ln( \rd ) - \mu}{\sqrt{2} \sigma}
 \right]^2
\right),& 
a>0,
\end{eqnarray}
will be employed to characterize materials' polydispersity.  The
parameters $\mu$ and $\sigma$ are provided by~\cite[Eq.~(17)]{Bohm}
\begin{eqnarray}
\mu 
= 
\ln( \rd_{50} ),
&
\displaystyle
\sigma
=
\frac{1}{1.2816}
\ln 
\left(
  \frac{S + \sqrt{S^2 + 4}}{4} 
\right),
&
S 
=
\frac{\rd_{90} - \rd_{10}}{\rd_{50}},
\end{eqnarray}
where $\rd_x$ denotes the $x$-th percentile of the particle
radii and $S$ is the span of the size distribution.

\section{Mori-Tanaka estimates - example results}\label{sec:MTexamples}

\subsection{Validation against available experimental data}\label{sec:Expvalid}
Two particular examples of real engineering materials are examined in
this section to show applicability of
Eq.~\eqref{eq:general_imperfect_system} and its extension for
polydisperse distribution of heterogeneities, Eq.~\eqref{eq:polydisp},
even when disregarding their actual shape and simply accepting a
spherical representation of the inclusions in the Mori-Tanaka
predictions. The results provided by these two equations are
corroborated by available experimental data.

\subsubsection{Random dispersion of copper particles in the Epoxy matrix}\label{sec:Example1}
In this first example we compare the \newtext{single-phase} Mori-Tanaka
predictions with the experimental results of de Araujo and
Rosenberg~\cite{Araujo} who measured the effective thermal conductivities of
systems consisting of random dispersion of metal particles in the epoxy matrix
for several values of the interfacial resistance arising due to acoustic
mismatch at the particle-matrix interface particularly for low temperatures
below $20$K. To enhance credibility of the Mori-Tanaka predictions we
focus on one particular system made from epoxy resin filled with copper particles also
examined in~\cite{Tor} in view of the three-point lower bound assuming a random
array of superconducting hard spheres (i.e. $\chi\phs{1}\rightarrow\infty$). It
has been shown, see~\cite{Araujo,Jackel:95:Cr}, that for metal-filled composites
with ratio $\alpha=\chi\phs{1}/\chi^{\mtrx}>10^2$ the macroscopic conductivity
does not depend on the thermal conductivity \newtext{of} the metallic filler,
but only on its volume fraction together with the properties of matrix and matrix-particle
interface. This becomes evident when rewriting
Eq.~\eqref{eq:general_imperfect_system}$_1$ in terms of $\alpha$
\begin{equation}\label{eq:eff_cond_alpha}
\frac{\KH}{\Km} 
= 
1
+
\frac{3\vfrac\phs{1}\left[\alpha-\left(1+R\right)\right]} {\vfrac^{\mtrx}\left[\alpha+2\left(1+R\right) \right]+3\vfrac\phs{1}\left(1+R\right)},
\end{equation}
where we introduced the dimensionless quantity $R$ adopted in~\cite{Tor}
\begin{equation}\label{eq:tor}
R
=
\frac{\chi\phs{1}}{ka}
=
\frac{\alpha\Km}{ka},
\end{equation}
\newtext{where $a$ is the sphere radius. It follows from
Eq.~\eqref{eq:eff_cond_alpha} that for the inclusion size equal to the Kapitza
radius~\cite{Every:1992:EPS}
\begin{equation}
a_\mathrm{K}
=
\frac{\alpha}{\alpha - 1}
\frac{\Km}{k}
\approx 
\frac{\Km}{k}
\end{equation}
the effective conductivity equals to that of the matrix, thus the effect of
inclusions becomes completely shielded by the interface. For $a < a_K$, the
overall properties are dominated by interfaces and the effective conductivity
decreases with increasing $\vfrac\phs{1}$ even when $\alpha > 1$. For $a > a_K$,
the bulk properties of phases become dominant; see also~\cite{Every:1992:EPS}
for further discussion.}

In addition to experimental measurements, we also present a comparison with the
Torquato-Rintoul three-point lower bound in resistance case \newtext{derived
in}~\cite[Eqs.~(8)]{Tor}, \newtext{evaluated for the statistical parameter
$\zeta(c\phs{1})$ given for the model of impenetrable spheres in~\cite[Table II
(simulations)]{Miller:1990:ECH}}. \newtext{Note that the interfacial properties
can be estimated by measuring the ratio of temperature jump to the applied heat
flux across a thin bi-material layer, by acoustic mismatch model~\cite{AlSi} or,
indirectly, from an inverse approach as discussed below.} The results are
presented for two different temperatures $\theta=4$K \newtext{($a=14.8
a_\mathrm{K}$)} and $\theta=3$K \newtext{($a=4.93 a_\mathrm{K}$)}.

\begin{figure}[hbtp]
\begin{center}
\begin{tabular}{c@{\hspace{2mm}}c@{\hspace{2mm}}c}
\includegraphics*[height=46mm]{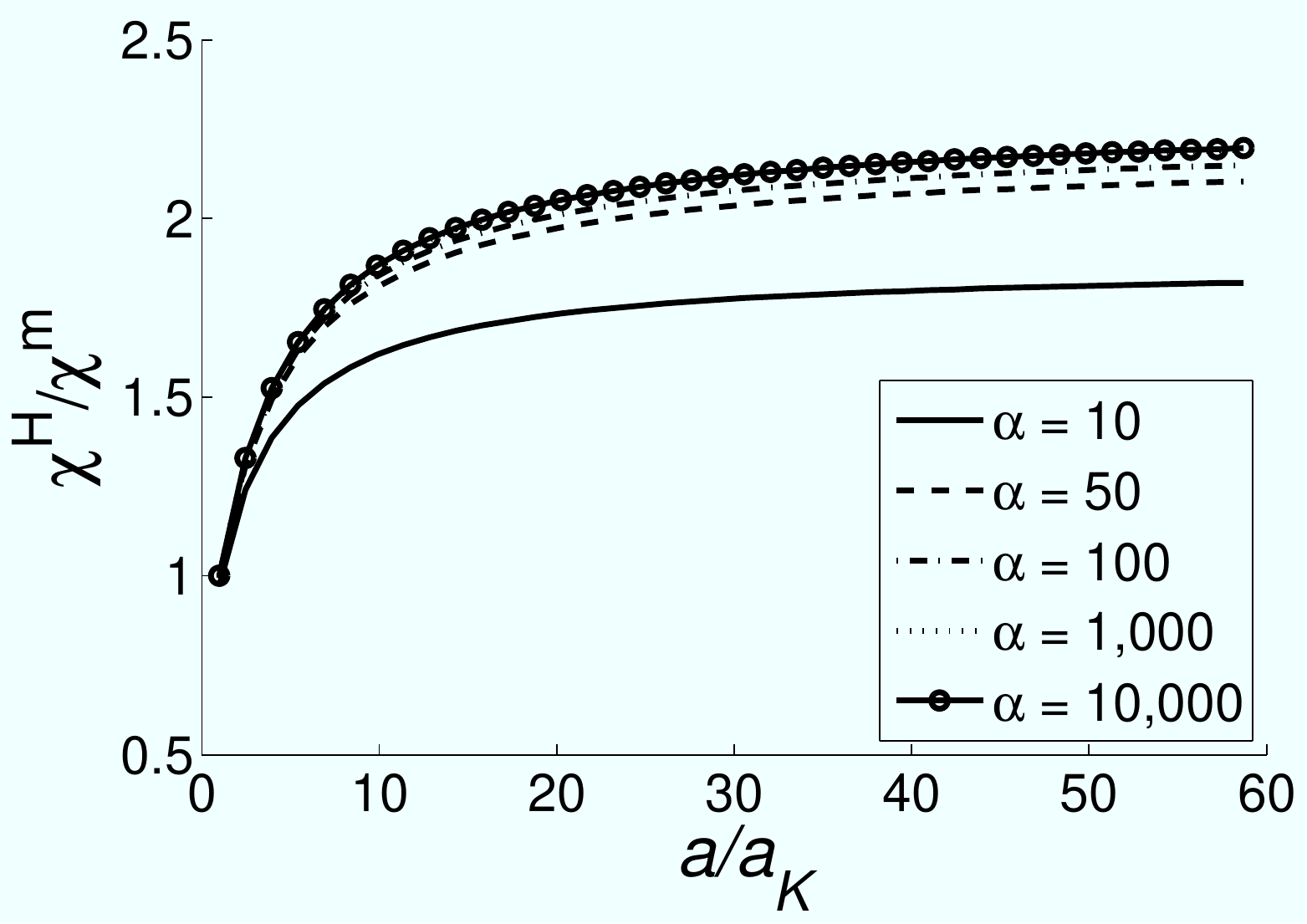}&
\includegraphics*[height=46mm]{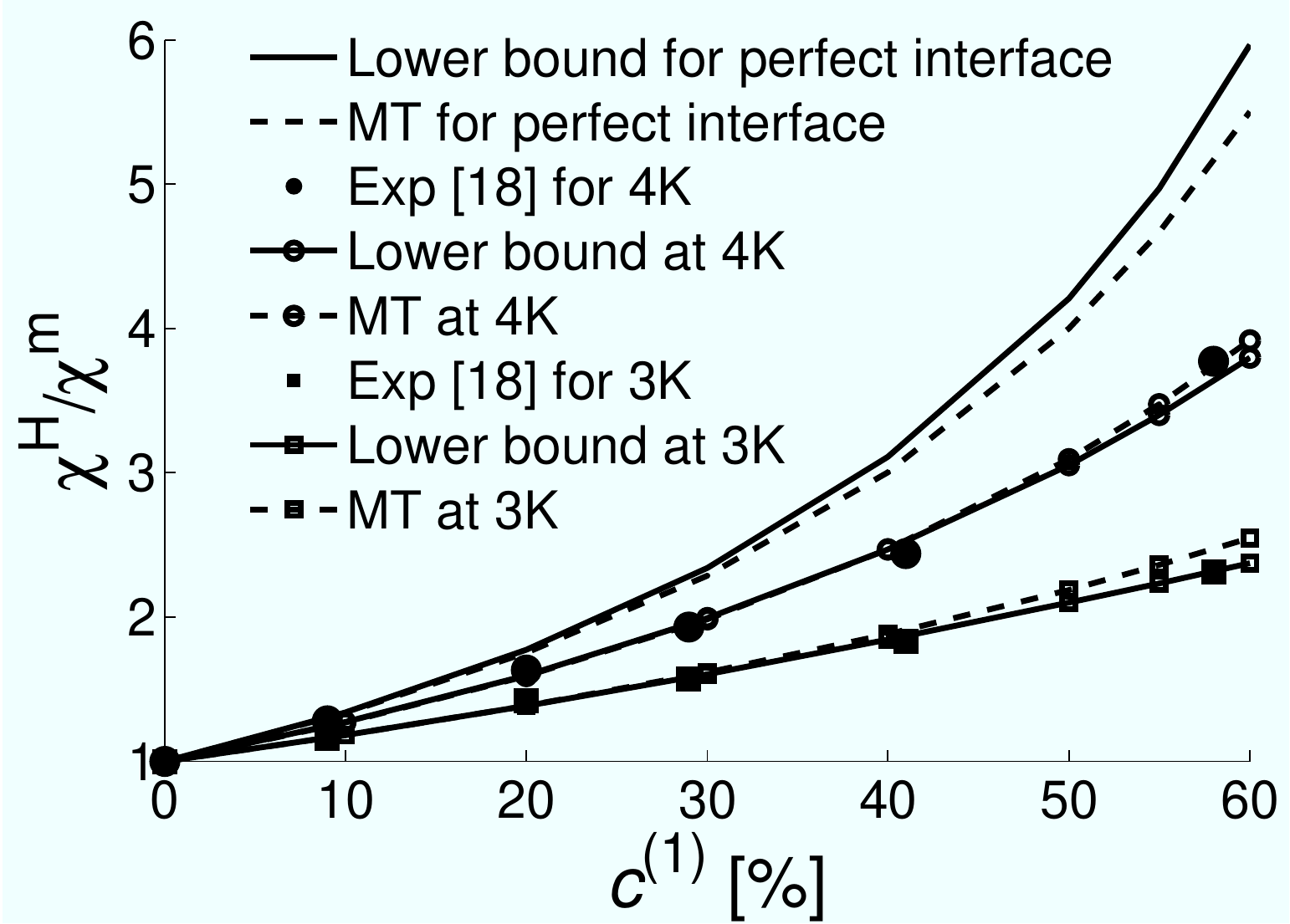}&
\includegraphics*[height=46mm]{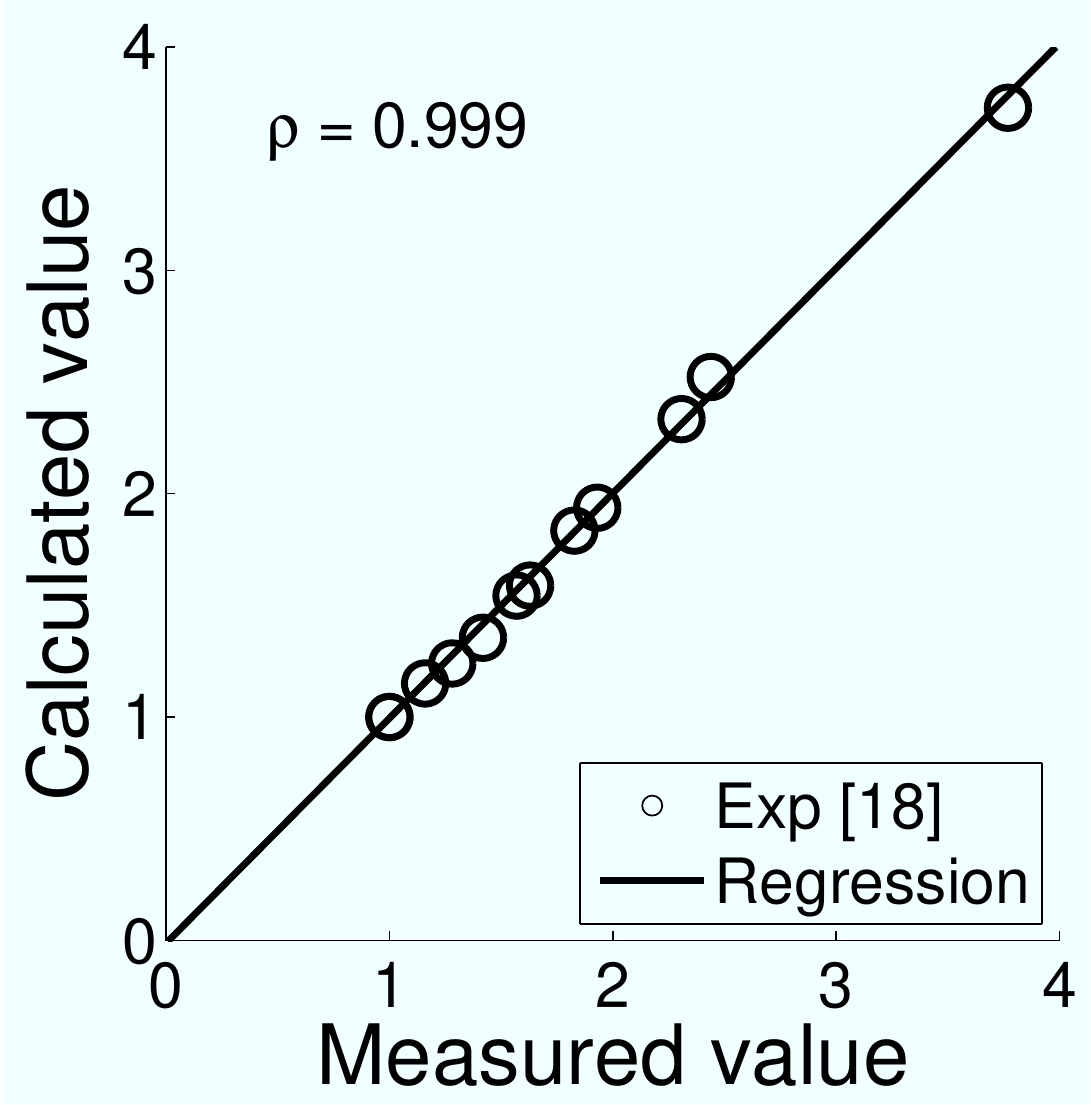}\\
(a)&(b)&(c)
\end{tabular}
\end{center}
\caption{Evolution of the \newtext{normalized effective} thermal conductivity
\newtext{$\KH/\Km$} as a function of \newtext{a) phase contrast $\alpha$ and
relative particle radius $a/a_\mathrm{K}$~($\vfrac\phs{1}=30\%$) and b)} volume
fraction \newtext{$\vfrac\phs{1}$} of copper particles, \newtext{c)}~correlation
of measured and calculated values; \newtext{$\rho$ is the correlation coefficient}.}\label{T_grafy}
\end{figure}
The influence of parameter $\alpha$ together with expected particle size
dependence, now hidden in parameter $R$ through Eq.~\eqref{eq:tor}$_1$, is
evident from~\figref{T_grafy}(a) plotted for $\theta=4$K
(\newtext{$a_\mathrm{K}=3.38~\mu$m}). Clearly, the Mori-Tanaka predictions
confirm negligible influence of $\chi\phs{1}$ observed for particulate
composites with $\chi\phs{1}\gg\chi^{\mtrx}$ ($\alpha>10^2$) as well as
decreasing trend for $\KH$ with decreasing particle size caused by imperfect
thermal contact. \figref{T_grafy}(b) then displays evolution of normalized
effective thermal conductivity as a function of volume fraction of copper
particles. Predictive capability of the Mori-Tanaka method is supported here by
a very good agreement with both experimental measurements and the
Torquato-Rintoul bounds~\cite{Tor}. Finally, \figref{T_grafy}(c) shows
correlation between theoretical (MT predictions) and experimental results. The
solid line was derived by linear regression of measured and calculated effective
conductivities using the least square method. Another indication of the quality
of numerical predictions can be presented in the form of Pearson's correlation
coefficient written as
\begin{equation}
\rho=\frac{n \langle{xy}\rangle - \langle{x}\rangle\langle{y}\rangle}{\sqrt{n\langle{x^2}\rangle - \langle{x}\rangle^2}\sqrt{n \langle{y^2}\rangle - \langle{y}\rangle^2}},
\end{equation}
where $\langle{a}\rangle=\sum_{i=1}^n a_i$, $n$ is the number of
measurements, $x$ stands for the experimental and $y$ for the
corresponding theoretical values.  Note that for $\rho=1$ the
correspondence is exact. In this case the correlation coefficient
equals to 0.999 suggesting almost perfect match between measured and
predicted values, also evident from graphical presentation.

\subsubsection{Al/SiC composite}\label{sec:Example2}
In~\cite{AlSi} the authors studied the effect of imperfect thermal
contact on the macroscopic response of Al/SiC porous
composites. The paper presents the results of a thorough experimental
investigation and traces of an inverse approach in material
mechanics for inferring material properties of unknown components of
the composite by matching numerical and experimental results. This
approach was first exploited to derive the matrix thermal conductivity
from known electrical conductivity of the composite. Next, the
Hasselman and Johnson model~\cite{Hasselman} was employed under the
premise of random distribution of spherical particles of identical
size to estimate the particle thermal conductivity and interfacial
thermal conductance for pore-free specimens and subsequently utilized
in the two-step application of the Hasselman-Johnson model to address
the influence of pores. Note that the material data used in these
predictions can be thought as optimal with respect to the adopted
Hasselman-Johnson model.

Hereinafter, we compare the results presented in~\cite{AlSi} for seven
specimens with pore-free matrix. In addition, we take advantage of
available characteristics of the SiC particles, the span $S$ and the
50-th percentile $a_{50}$, to extend the analysis to polydisperse
systems as presented in~\secref{sec:PolydisperseSystems}. The input
material data are listed in~\tabref{tab-AlSiC-mat}. 

\begin{table}[ht]
\caption{Material properties~\cite{AlSi}.}
\label{tab-AlSiC-mat}
\centering
\begin{tabular}{c|c|c}\hline
Al matrix & SiC particles & Interface  \\
\hline
\multicolumn{2}{c|}{[$\mathrm{Wm^{-1}K^{-1}}$]}&[$\mathrm{Wm^{-2}K^{-1}}$]\\
\hline
187 & 252.5 &72.5$\times 10^6$\\\hline
\end{tabular}
\end{table}

\begin{figure}[hbtp]
\begin{center}
\begin{tabular}{c@{\hspace{0mm}}c@{\hspace{0mm}}c}
\includegraphics*[height=46mm]{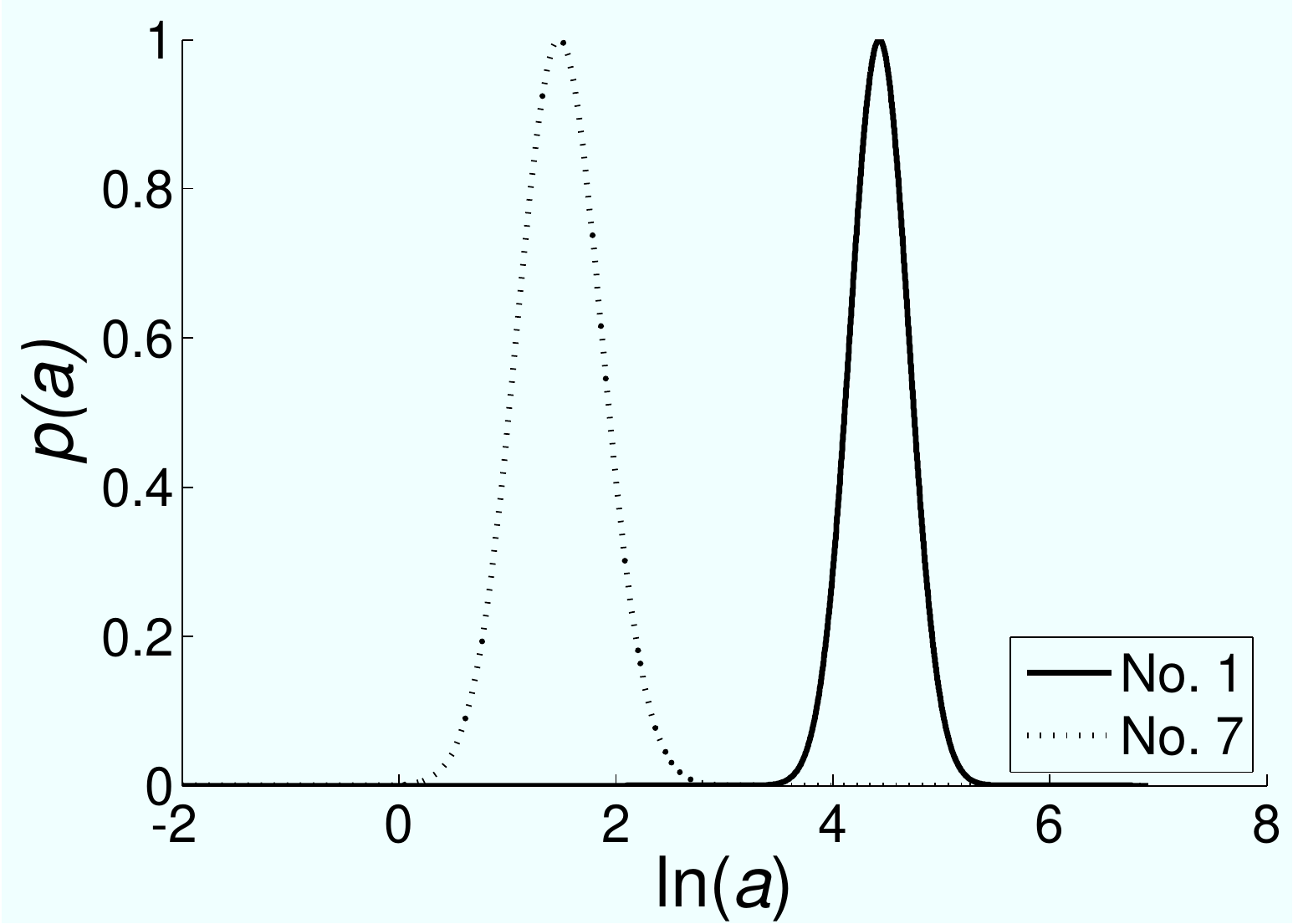}&
\includegraphics*[height=46mm]{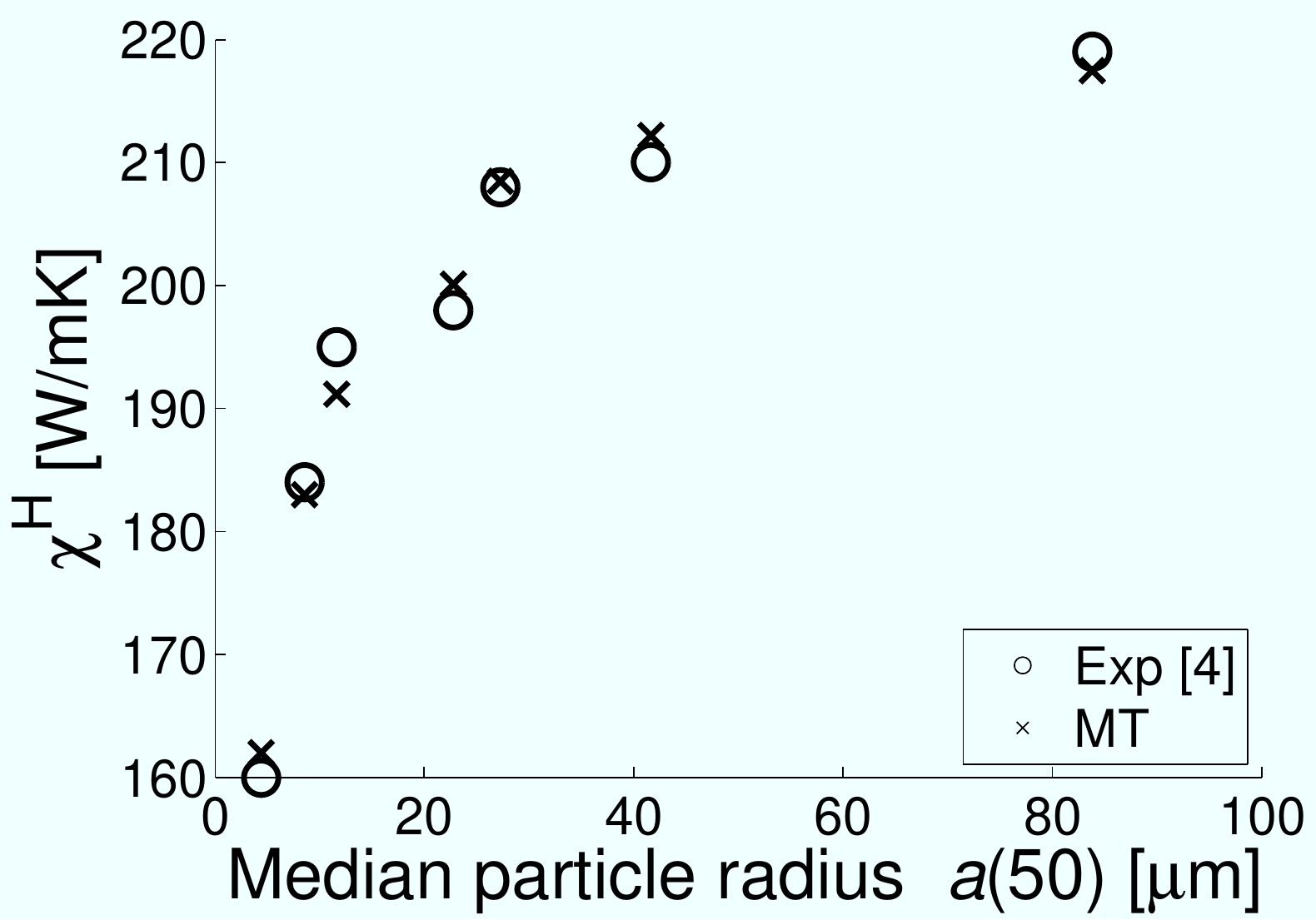}&
\includegraphics*[height=46mm]{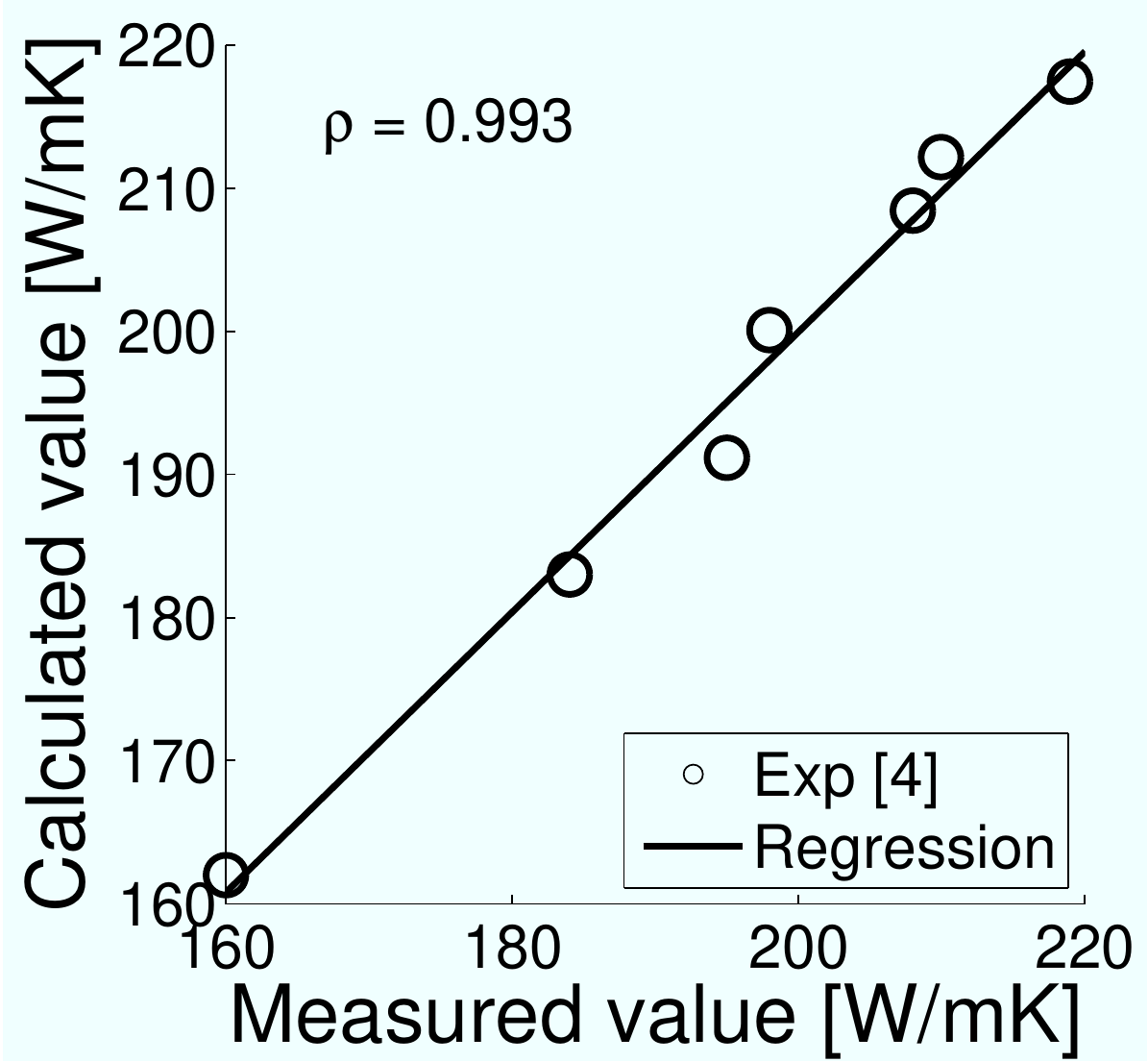}\\
(a)&(b)&(c)
\end{tabular}
\end{center}
\caption{a) Examples of probability distribution
functions \newtext{of particle radii $a$} for specimens No. 1 and 7, b)
evolution of effective thermal conductivity \newtext{$\KH$} as a function of
\newtext{particle radius $a$}, c)~correlation of measured and calculated
values; \newtext{$\rho$ denotes the correlation coefficient}.}\label{S_grafy}
\end{figure}

The Mori-Tanaka predictions stored in~\tabref{tab-AlSiC-res} were
provided by Eq.~\eqref{eq:polydisp}. The integral~\eqref{eq:integ} was
evaluated such that the entire interval was split into 1,000 segments,
thus considering 1,000 different particle sizes of spherical
shape. Within each segment $s$ the probability function $p(a)$ was
approximated by a straight line and the volume fraction $c\phs{1}$ of a
given set of particles, given by the segment area, was then associated
in a logarithmic way with the mean radius $a\phs{1}$ of this set of
particles.  Standard trapezoidal integration rule was then used to sum
over all 1,000 segments. Examples of probability distribution functions
for specimens No. 1 and 7 are plotted in~\figref{S_grafy}.

\begin{table}[ht]
\caption{Characteristics of SiC particles (cf.~\cite[Table 1]{AlSi}),
  and comparison of effective thermal conductivity $\KH$
  [$\mathrm{Wm^{-1}K^{-1}}$] between experimental
  measurements~\cite{AlSi} and MT predictions Eq.~\eqref{eq:polydisp}.
}
\label{tab-AlSiC-res}
\centering
\begin{tabular}{ccccccc}\hline
Sample & \multicolumn{3}{c}{Radius $\mathrm{\left[\mu m\right]}$} & 
SiC &\multicolumn{2}{c}{Results}\\ \cline{2-4} \cline{6-7}
No. & $a_{10}$ & $a_{90}$ & $S$ & vol. & Exp. & MT \\ \hline
1 & 55 & 114.5 & 0.71 & 0.58 & 219 & 217.8 \\
2 & 23  & 65.5 & 1.02 & 0.58 & 210 & 212.3 \\
3 & 19.5  & 37.5  & 0.66 & 0.60 & 208 & 208.5 \\
4 & 11.5  & 25  & 0.79 & 0.59 & 198 & 199.9 \\
5 & 7  & 17  & 0.86 & 0.58 & 195 & 190.8 \\
6 & 5  & 12  & 0.82 & 0.55 & 184 & 182.5 \\
7 & 2.4 &  7 & 1.05 & 0.53 & 160 & 161.3 \\
\hline
\end{tabular}
\end{table}

Graphical representation of the results is further seen
in~\figref{S_grafy}(b),(c) confirming the sensitivity of the effective
thermal conductivity to the mean particle size distribution. Note that
individual points in~\figref{S_grafy}(b) correspond to slightly
different volume fraction, see~\tabref{tab-AlSiC-res}.

Almost negligible deviation from experimental results measured as
\begin{equation*}
E=\sqrt{\frac{\displaystyle{\sum_i}(\chi^{\sf exp}-\chi^{\sf
MT})^2}{\displaystyle{\sum_i}(\chi^{\sf MT})^2}}=1.1\%
\end{equation*}
is, however, not surprising owing to the used material parameters,
which were not measured but rather fitted to a micromechanical
model. Even when comparing the Pearson correlation coefficients, $0.98$
for the Hasselman-Johnson model and $0.993$ for polydisperse MT model,
the improvement when accounting for more accurate representation of
particle size distribution is marginal. This can be explained by a
very small variance associated with adopted distributions,
recall~\figref{S_grafy}(a).  Nevertheless, it is fair to point out
that unlike the Hasselman-Johnson model the Mori-Tanaka approach is
not limited to spherical particles providing the transformations given
by Eqs.~\eqref{eq:eff_isotropic_conductivity}
and~\eqref{eq:general_imperfect_system} are admissible. The influence
of shape of particles on the macroscopic response has been put
forward, e.g. by J\"{a}ckel in~\cite{Jackel:95:Cr}, and is numerically
investigated in the next section suggesting increasing thermal
conductivity of the composite with transition from spherical to
needle-like particles, the trend also observed
experimentally~\cite{Araujo,Jackel:95:Cr}.

\subsection{Verification against finite element simulations}\label{sec:Numverif}
It has been argued in the previous sections that even very limited
information about microstructure amounted to phase properties and
corresponding volume fractions might be sufficient to provide a
reasonable estimate of macroscopic response of various engineering
material systems generally classified as being macroscopically
isotropic. This naturally invites the assumption of spherical
representation of otherwise irregular heterogeneities. Although
supported by several practical examples discussed in the previous
section, we should expect and even identify, at least qualitatively,
limitations to such perception. In doing so, this section presents
numerical investigation of some specific issues such as the influence
of shape and size of inclusions or mismatch of phase material
properties on the predicted macroscopic response.

\newtext{All numerical results reported bellow are obtained in the
two-dimensional setting, hence the arguments presented in \secref{sec:MTthoery}
need to be translated to the planar case. In particular, inclusions are
modeled as elliptic cylinders of aspect ratio $\ar_2 = a_2 / a_1$ with $a_3
\rightarrow \infty$. The corresponding Eshelby-like tensor is given by
Eq.~\eqref{eq:eshelby_2D}, which leads to the two-dimensional thermal
concentration factor for circular inclusion in the form
\begin{eqnarray}\label{eq:2D_conc_factor}
\mCHi 
=
\CHi\cir
\I_{(2 \times 2)}
& \mbox{with} &
\CHi\cir 
=
\frac{2 \Km}{\Km + \Ki}.
\end{eqnarray}
The apparent conductivity due to random orientation is provided by 
\begin{equation}\label{eq:2D_orientation_averaging}
\widetilde{\K}\phs{s}
=
\frac{\oavg{\K\phs{s}\pCH{s}}_{2\mathrm{D}}}{\oavg{\pCH{s}}_{2\mathrm{D}}}. 
\end{equation}
where $\oavg{\bullet}_{2\mathrm{D}}$ stands for the orientation averaging for
the uniform distribution of the Euler angle $\phi$. Finally note that the
apparent conductivity due to imperfect interface is provided by the same
relation as in the three-dimensional setting, compare
Eq.~\eqref{eq:single_particle_imperfect} with Eq.~\eqref{eq:Keq_3D}.}

\begin{figure}[ht]
\begin{center}
\begin{tabular}{c@{\hspace{10mm}}c@{\hspace{10mm}}c}
\includegraphics*[width=40mm,keepaspectratio]{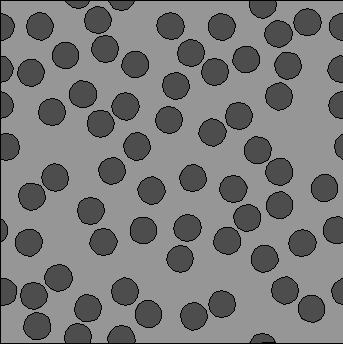}&
\includegraphics*[width=40mm,keepaspectratio]{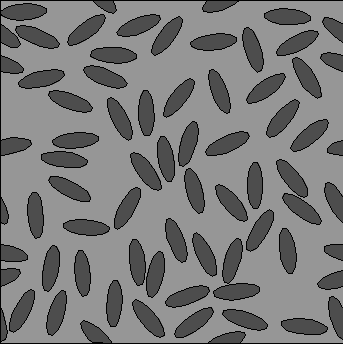}&
\includegraphics*[width=40mm,keepaspectratio]{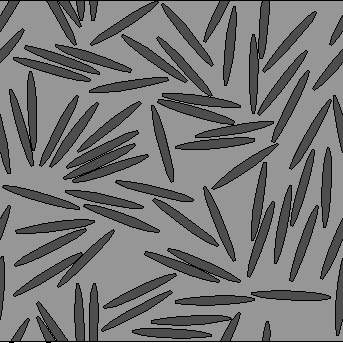}\\
(a)&(b)&(c)
\end{tabular}
\end{center}
\caption{Examples of random macroscopically isotropic microstructures:
a)~circular cylinders~($\ar_2 = 1$), 
b)~elliptical cylinders with aspect ratio \newtext{$\ar_2 = 3$}, 
c)~elliptical cylinders with aspect ratio \newtext{$\ar_2 = 9$}.}
\label{fig:shape-geom}
\end{figure}

\begin{figure}[ht]
\begin{center}
\begin{tabular}{c@{\hspace{0mm}}c@{\hspace{0mm}}c}
\includegraphics*[width=.33\textwidth]{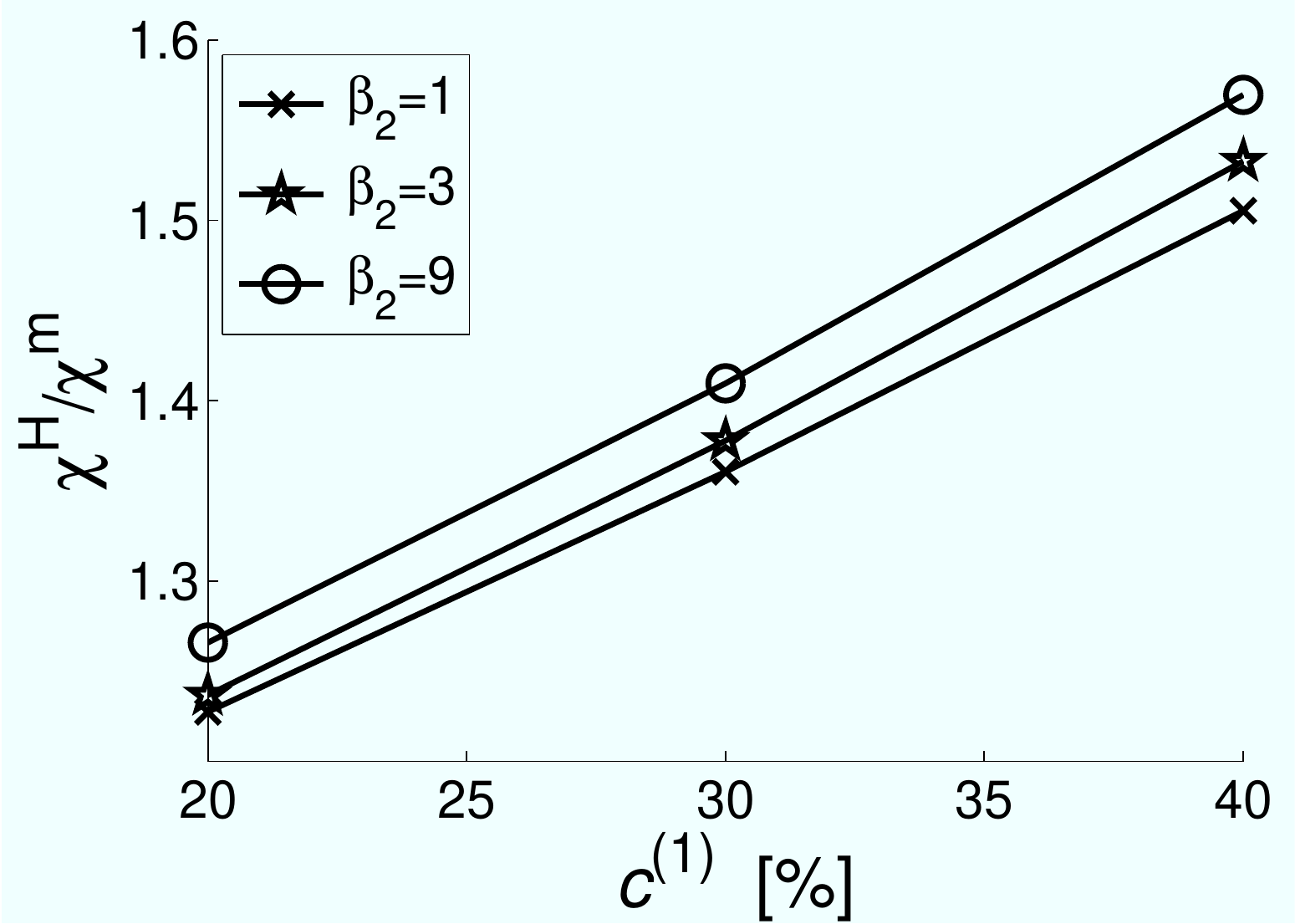}&
\includegraphics*[width=.33\textwidth]{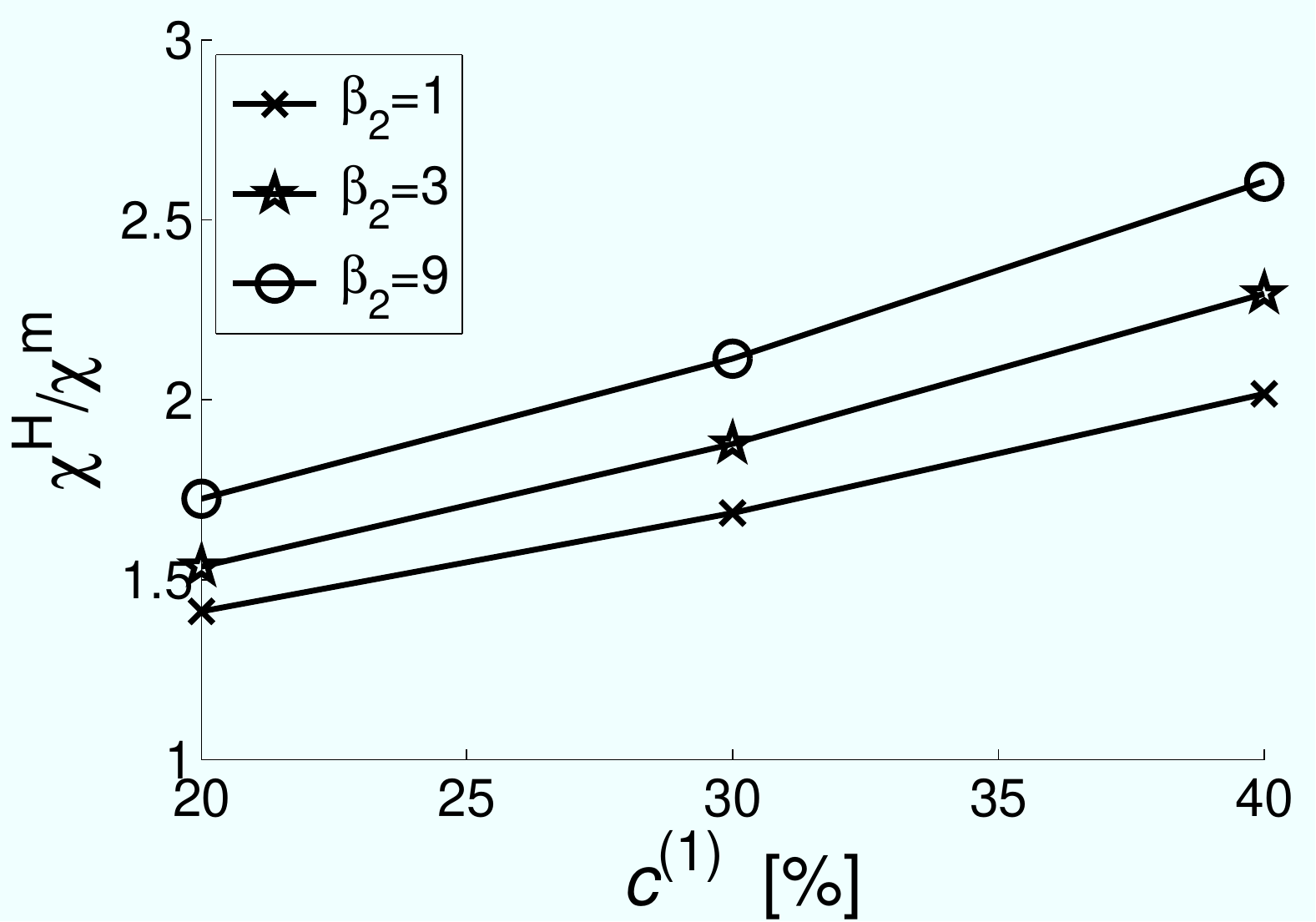}&
\includegraphics*[width=.33\textwidth]{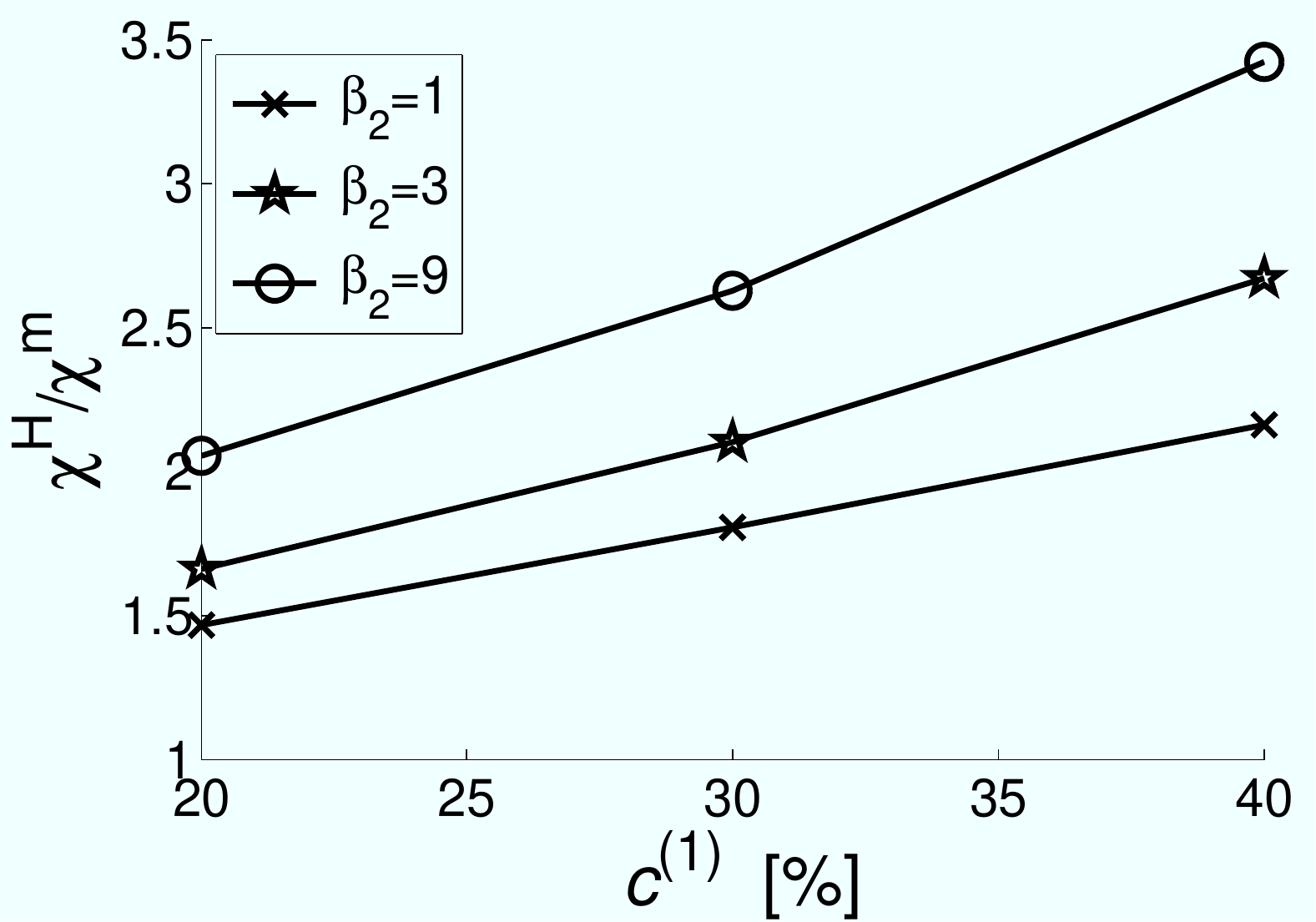}\\
(a)&(b)&(c)
\end{tabular}
\end{center}
\caption{Variation of \newtext{the normalized} effective conductivity
\newtext{$\KH/\Km$} for three microstructures in~\figref{fig:shape-geom}
\newtext{with perfect interfaces, determined by periodic FEM homogenization for
phase contrast a) $\alpha = 3$, b) $\alpha = 10$ and c)~$\alpha = 20$; $\ar_2$
denotes the inclusion aspect ratio and $\vfrac\phs{1}$ the inclusion
volume fraction.}}
\label{fig:shape-res}
\end{figure}

Three particular representatives, generated such as to approximately resemble
the real microstructures in~\figref{fig:eng-mat}, appear
in~\figref{fig:shape-geom}. To comply with general assumptions put forward in
the previous sections, we consider locally isotropic phases with variable
contrast of material properties. Additionally, we assume the above
microstructures being periodic and adopt the classical first-order
homogenization strategy, see e.g.~\cite{Michel:1999:EPC,Zeman:2001:EPG}, to
provide estimates of the macroscopic response. The results are plotted
in~\figref{fig:shape-res}.

These results clearly indicate not only the influence of the shape of inclusions
on the macroscopic response but also a strong dependence of these predictions on
the contrast of material properties of individual phases. Thus drawing from the
plots presented in~\figref{fig:shape-res}(a) one may suggest that the proposed
\newtext{circular} representation of generally non-\newtext{circular}
heterogeneities is still acceptable when their shapes only moderately deviate
from a \newtext{circle} and when the mismatch of phase properties is not too
severe, which certainly is the case of a number of real materials as demonstrated in the
previous section. This expectation is quite important particularly when dealing
with imperfect \newtext{thermal} contact in which case only spherical
\newtext{and circular} inclusions can be \newtext{easily} handled analytically.

\begin{figure}[ht]
\begin{center}
\begin{tabular}{c@{\hspace{0mm}}c@{\hspace{0mm}}c}
\includegraphics*[width=.33\textwidth]{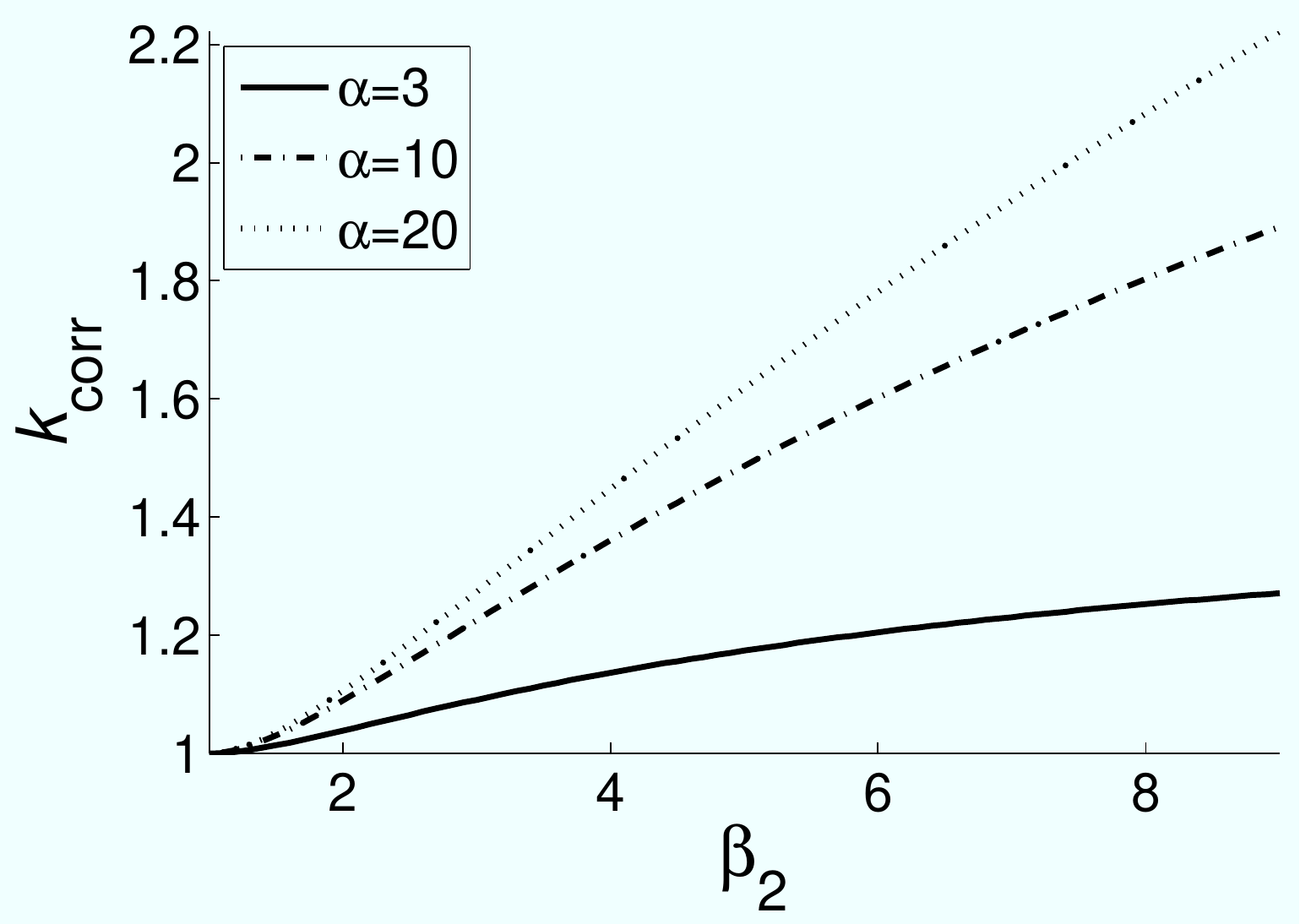}&
\includegraphics*[width=.33\textwidth]{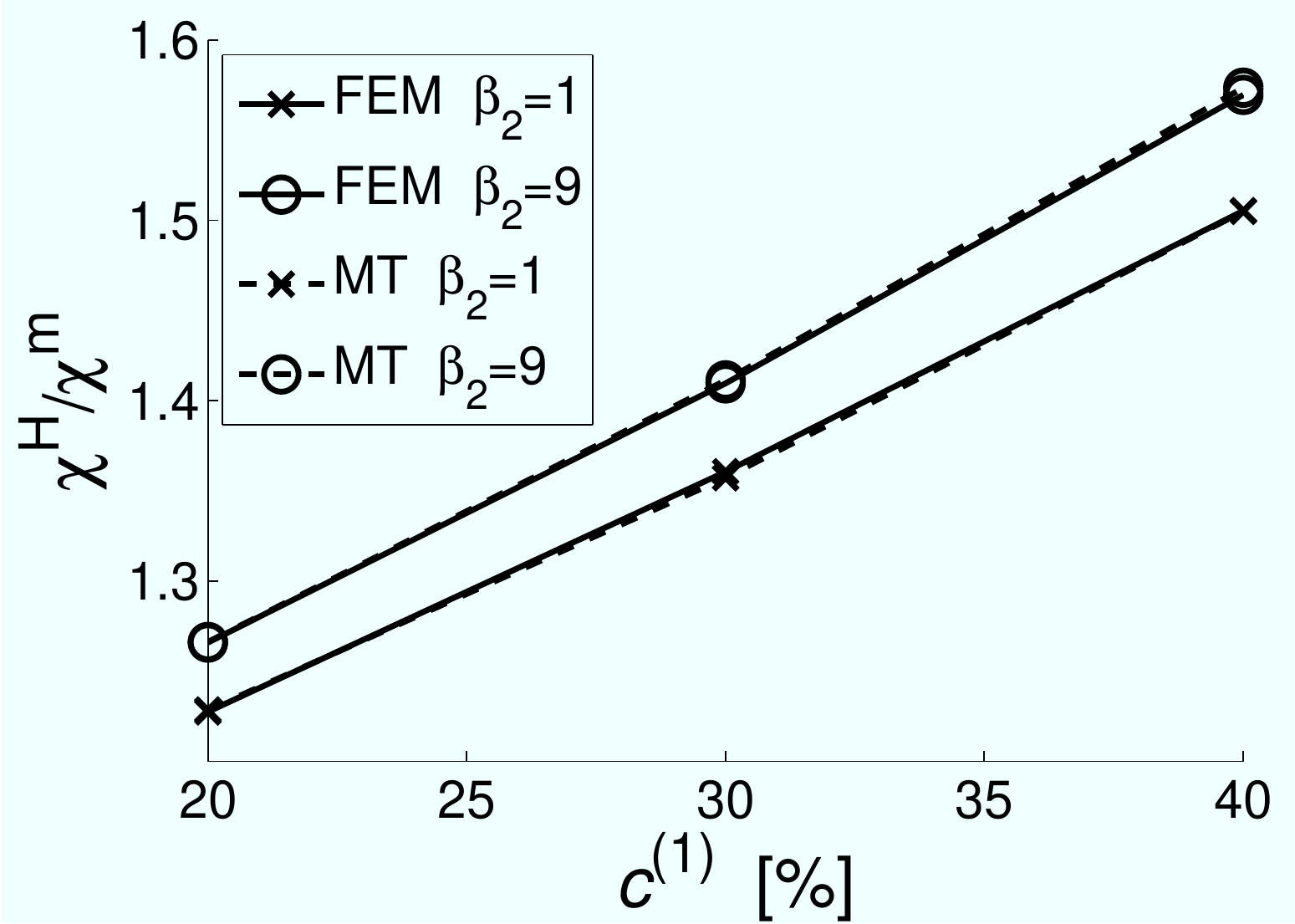}&
\includegraphics*[width=.33\textwidth]{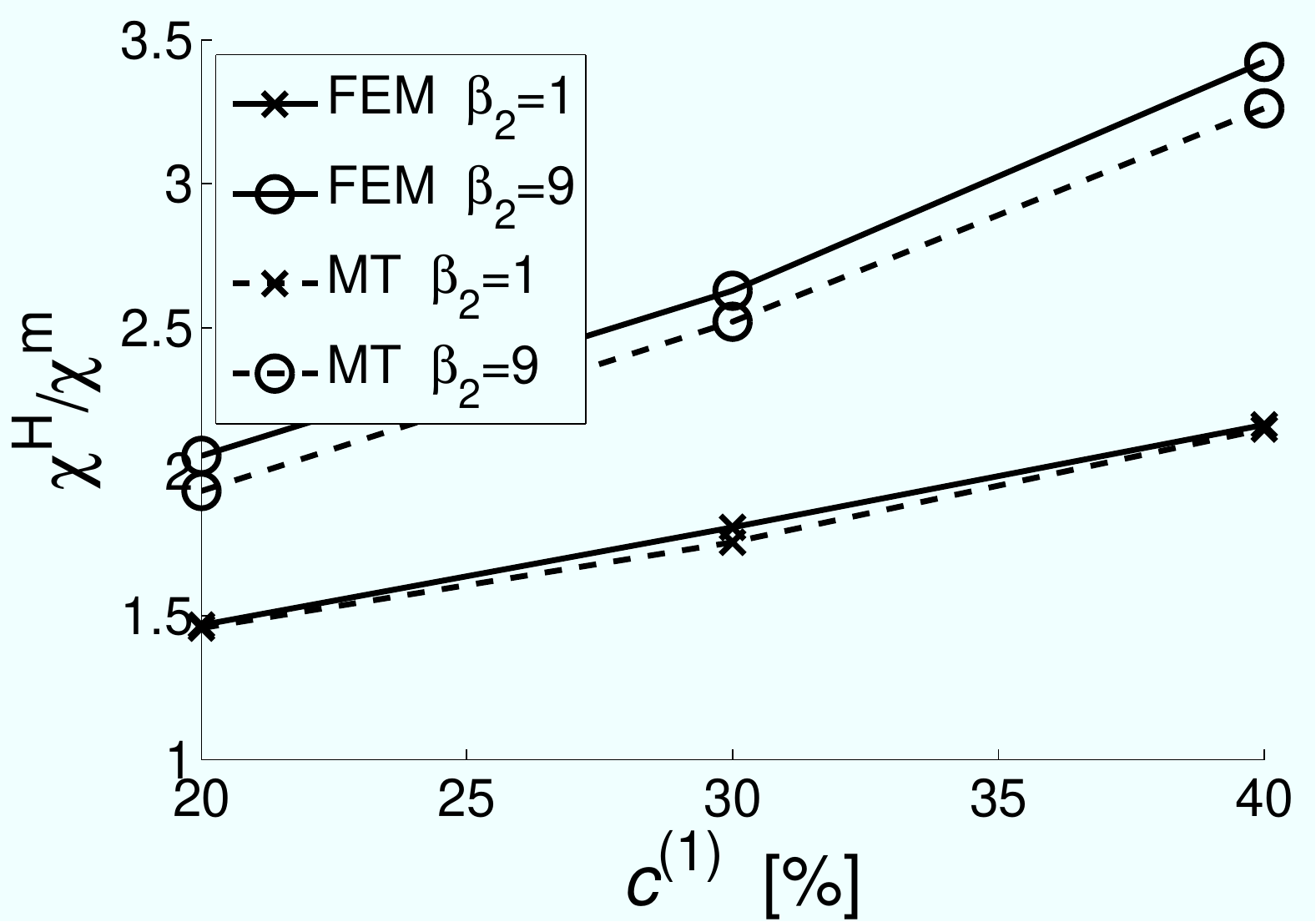}\\
(a)&(b)&(c)
\end{tabular}
\end{center}
\caption{a) Evolution of correction factor $k_{\sf corr}$ \newtext{for systems
with perfect interfaces as a function of the ratio of semi-axes
\newtext{$\ar_2$} and
variation of normalized} effective conductivity \newtext{$\KH / \Km$} obtained
by FEM and MT with modified conductivity $\widetilde{\chi}\phs{1}$ for phase
contrast b) $\alpha = 3$ and c)~$\alpha = 20$.}
\label{fig:kcorr}
\end{figure}

If the \newtext{circular} approximation of heterogeneities is no
longer acceptable or the contrast of phase properties is excessive, one needs to
look for more details about microstructure. In such \newtext{a} case, even
two-dimensional images of real systems, at present almost standard input for any material based
analysis, may play an important role in assessing better approximations of
shapes, say \newtext{elliptical}, of these heterogeneities, see
e.g.~\cite{Tsukrov:MAMS:2005}. Then, being given the \newtext{elliptical} shape
of the inclusion allows us to appropriately modify its material data, recall
\newtext{Eq.}~\eqref{eq:2D_orientation_averaging}, and define a certain
indicator of the real microstructure $k_{\sf corr}$, 
e.g. as a ratio of the modified and original partial temperature gradient concentration factors 
\begin{equation}
k_{\sf corr}=\frac{\widetilde{T}\phs{1}_\mathrm{cir}}{{T}\phs{1}_\mathrm{cir}},
\tag{39}
\end{equation}
now determined for circular inclusions.

Variation of this parameter as a function of the shape of inclusion is seen
in~\figref{fig:kcorr}(a) further confirming quite strong influence of the phase
properties mismatch.\footnote{\newtext{Note that the parameter $k_{\sf corr}$ is
determined for two dimensional systems and thus is not applicable to ellipsoidal
inclusions of the same aspect ratio.}} The modified conductivity when introduced
successively into \newtext{Eq.}~\eqref{eq:2D_conc_factor}$_2$ and
Eq.~\eqref{eq:eff_isotropic_conductivity}$_1$ then renders the estimate of
effective conductivity almost identical to actual microstructure with
non-\newtext{circular} inclusions as evident from plots
in~\figref{fig:kcorr}(b),(c). Note that only the first and the third
microstructure in~\figref{fig:shape-geom} were examined to first confirm that
the Mori-Tanaka method is indeed well suited for statistically isotropic random
microstructures and second to promote applicability of this simple
transformation from \newtext{elliptical} to \newtext{circular} representations
even for shapes markedly distinct from \newtext{circles}. Small \newtext{but
evident} deviation of the results observed in \figref{fig:shape-geom} for the
third needle-like microstructure and large mismatch of conductivities of the
inclusion and matrix equal to~$20$ can be attributed to the finite size,
although infinite in the sense of periodicity, of the representative model not
large enough to yield statistically isotropic microstructure.

\begin{figure}[ht]
\begin{center}
\begin{tabular}{c@{\hspace{10mm}}c@{\hspace{10mm}}c}
\includegraphics*[width=40mm,keepaspectratio]{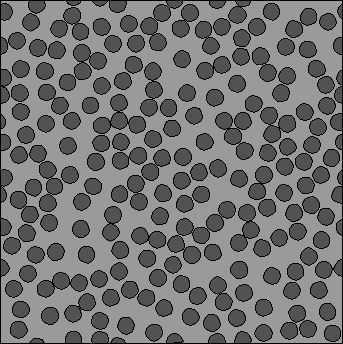}&
\includegraphics*[width=40mm,keepaspectratio]{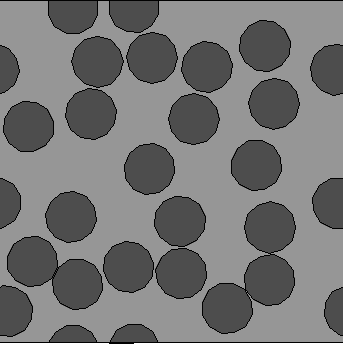}&
\includegraphics*[width=40mm,keepaspectratio]{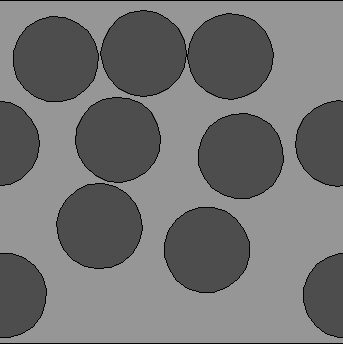}\\
(a)&(b)&(c)
\end{tabular}
\end{center}
\caption{Statistically isotropic distribution of circular cylinders with variable radius of their cross-section.}
\label{fig:circle-size-geom}
\end{figure}

The second set of numerical simulations addresses the theoretically
derived dependence of macroscopic predictions on the size of
inclusions in cases with imperfect interfaces generating jumps in the
local temperature field. \newtext{For simplicity, } we limit our attention to
statistically isotropic distributions of circular cylinders with a
radius varying from sample to sample. Three such microstructures are
shown in~\figref{fig:circle-size-geom}.  Note that the same volume
fraction and the same number of inclusions was maintained in all
simulations. Zero thickness interface elements were introduced
to account for imperfect thermal contact.

\begin{figure}[ht]
\begin{center}
\begin{tabular}{c@{\hspace{10mm}}c}
\includegraphics*[height=45mm]{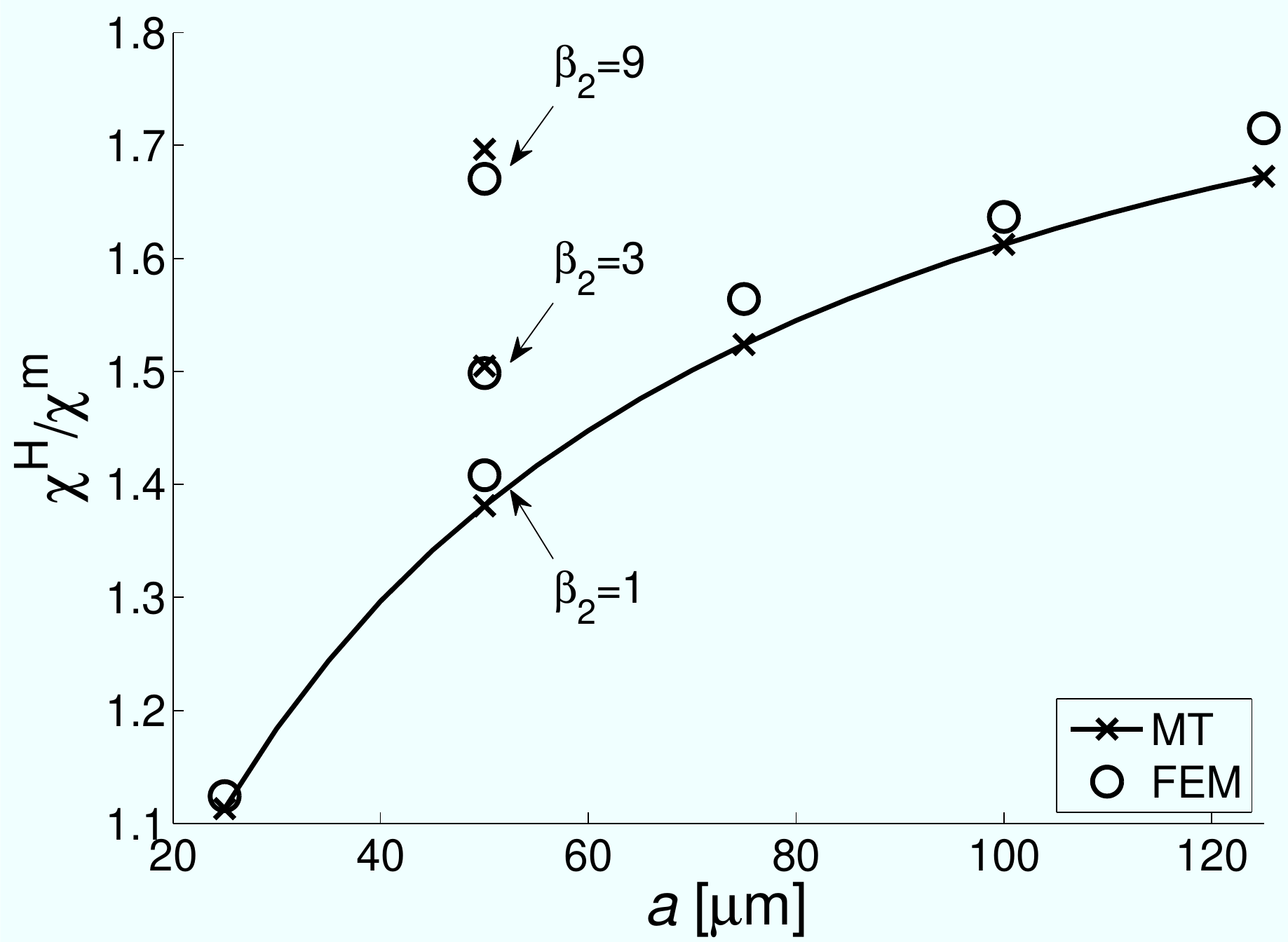}&
\includegraphics*[height=45mm]{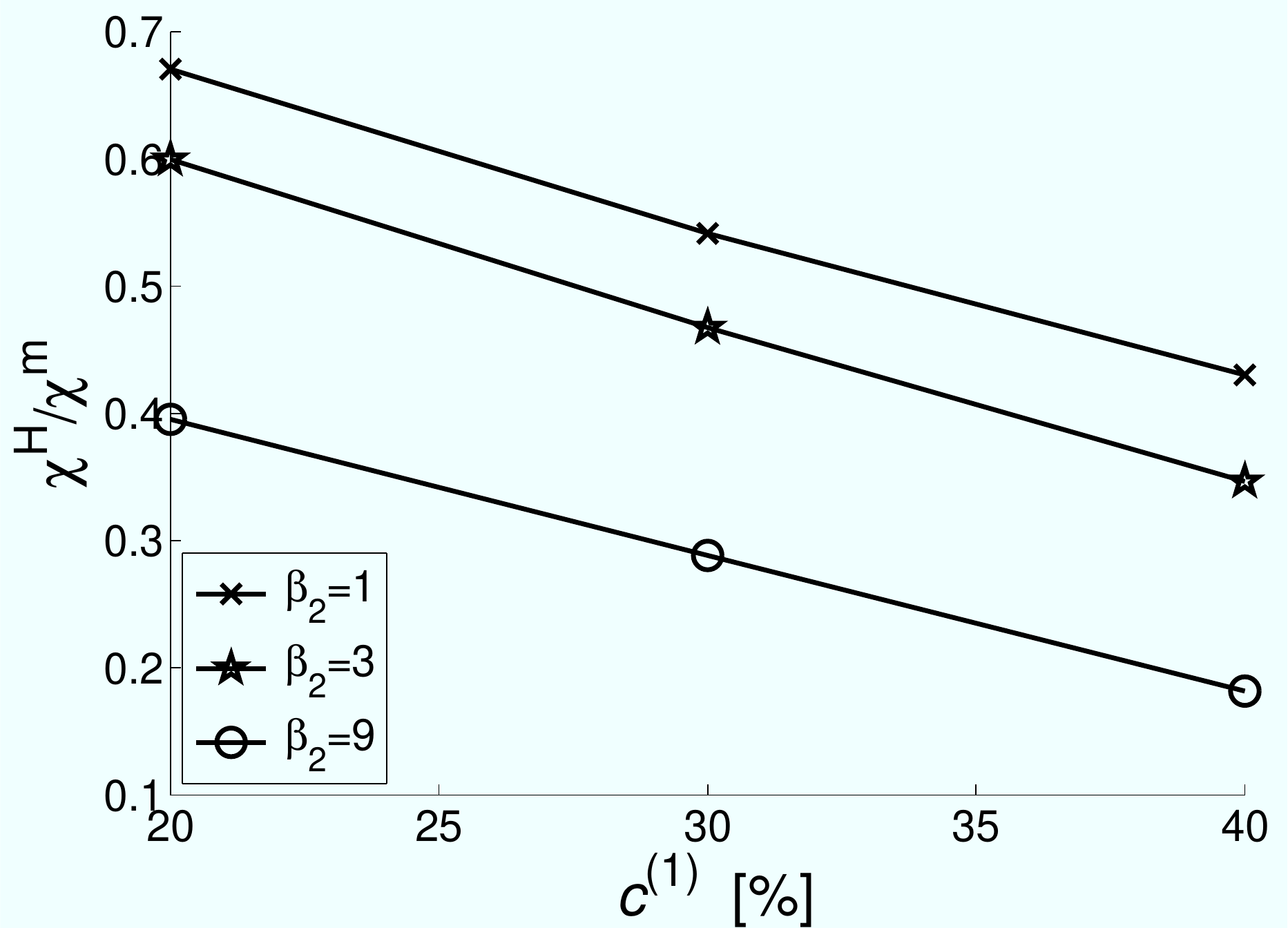}\\
(a)&(b)
\end{tabular}
\end{center}
\caption{a) Variation of \newtext{normalized} effective conductivity
\newtext{$\KH / \Km$} for a system \newtext{with imperfect interface~($k=60
\times 10^6~\mathrm{Wm^{-2}K^{-1}}$ and $\K\phs{1} =
100~\mathrm{Wm^{-1}K^{-1}}$, $\Km = 10~\mathrm{Wm^{-1}K^{-1}}$ and
$\vfrac\phs{1}=40\%$)} as a function of inclusion radius $a$ and b) variation of
effective conductivity for systems weakened by cylindrical voids \newtext{as a
function of inclusion volume fraction $\vfrac\phs{1}$; $\ar_2$ denotes inclusion
aspect ratio}.}
\label{fig:circle-size-res}
\end{figure}

The relevant results appear in~\figref{fig:circle-size-res}(a). Both
the results found from finite element simulations and corresponding
Mori-Tanaka predictions are displayed to clearly identify the
\newtext{mentioned} size dependence. Proper modifications in the sense of
Eq.~\eqref{eq:2D_orientation_averaging}
now become even more important as indicated by the results generated
for elliptical microstructures from~\figref{fig:shape-geom} with
cross-sectional area \newtext{equal} to the area of the circle.  These are
indicated by x-symbol and the ratio of semi-axes of the corresponding
elliptical cylinder. The Mori-Tanaka
estimates found from the application of
Eqs.~\eqref{eq:2D_orientation_averaging}
and \eqref{eq:single_particle_imperfect} are reasonably close further
supporting the proposed approach for the modeling of real materials with an imperfect thermal
contact. Intuitively, it can be expected that the value of interfacial
conductance $k$ may also show some effect as to the estimates of
effective conductivities for non-\newtext{circular} inclusions. This
notion is supported by the results presented in~\figref{fig:circle-size-res}(b)
showing variation of effective conductivity of an isotropic matrix
weakened by voids with a very low conductivity. Clearly, the influence
of shape of the inclusions is quite pronounced.

\section{Conclusions}\label{sec:concl}

The Mori-Tanaka micromechanical model has been often the primary
choice among engineers to provide quick estimates of the macroscopic
response of generally random composites. Motivated by an early
theoretical as well as experimental works on this subject, the
Mori-Tanaka method was examined here in the light of the solution of a
linear steady state heat conduction problem allowing us to estimate
the effective thermal conductivity of variety of real engineering
materials experiencing an imperfect thermal contact along the
constituents interfaces.

Adhering to the only limitation, an assumption of macroscopically isotropic
composite, it was shown that the \newtext{method originally proposed by B\"{o}hm
and Nogales~\cite{Bohm}} for a spherical representation of particles still
applies even to non-spherical particles providing their shape can be suitably
quantified, e.g. by an ellipsoidal inclusion. In this particular case the
Mori-Tanaka predictions were \newtext{partially} corroborated by
\newtext{two-dimensional} numerical simulations confirming experimentally
observed considerable sensitivity of macroscopic conductivities to the shape of
particles.

The fact that for composites with imperfect thermal contact the
macroscopic predictions depend on particle size can be effectively
handled by introducing the particle size probability density function
directly into the Mori-Tanaka estimates. Although not confirmed for
material systems studied in the paper, this may considerably improve
final predictions especially for grading curves showing significant
standard deviation of particle sizes from its mean value. This is
particularly appealing, since grading curves are one of the few
information supplied by the manufacturer.

To conclude, it is interesting to point out that there exist many
material systems that can be handled very effectively with simple
micromechanical models with no need for laborious finite element
simulations of certain representative volumes of real microstructures.

\section*{Acknowledgments} 
\newtext{The authors are thankful to two
anonymous referees for their constructive remarks and suggestions on the
original version of the manuscript.} The financial support provided by the
GA\v{C}R grants No.~106/08/1379 and P105/11/0224 and partially also by the
research project CEZ~MSM~6840770003 is gratefully acknowledged.

\appendix\label{sec:appen}

\section{Eshelby-like tensor}\label{sec:appa}
The Eshelby-like tensor for the solution of thermal conductivity
problem was introduced by Hatta and Taya in~\cite{EIM}. For an
ellipsoidal inclusion with semi-axes $a_1,a_2,a_3$ found in an
isotropic matrix it receives the form
\begin{eqnarray}\label{eq:S}
S_{ij} &=&
\frac{a_1a_2a_3}{4}\frac{\partial}{\partial x_i \partial x_j}
\int_0^\infty\left(\frac{x_1^2}{a_1^2+s}+
\frac{x_2^2}{a_2^2+s}+ \frac{x_3^2}{a_3^2+s}\right)
\frac{1}{\Delta
s}\mathrm{d}s,  \\
\Delta s &=&
\sqrt{\left(a_1^2+s\right)\left(a_2^2+s\right)\left(a_3^2+s\right)}.
\end{eqnarray}
Closed form solutions of integral~\eqref{eq:S} for some special cases
of ellipsoidal shapes of the inclusion can be found in~\cite{EIM}. For
a general ellipsoid the solution was introduced by Chen and Yang in
\cite{Chen:1995:AM}. For circular and spherical shapes needed in the
present study the $\Et$ tensor reads
\begin{itemize}
\item Sphere $\left(a_1=a_2=a_3\right)$
\begin{equation}
S_{11}=S_{22}=S_{33} = \frac{1}{3} \quad\mbox{and}\quad S_{ij}=0 \quad\mbox{for}\quad i\neq j.
\end{equation}
\item Elliptic cylinder $\left(a_3\rightarrow\infty\right)$
\begin{equation}\label{eq:eshelby_2D}
S_{11}=\frac{a_2}{a_1+a_2},\quad S_{22}=\frac{a_1}{a_1+a_2},\quad S_{33} = 0 \quad\mbox{and}\quad S_{ij}=0 \quad\mbox{for}\quad i\neq j.
\end{equation}
\end{itemize}

\section{Single spherical homogeneity with imperfect interface}\label{sec:appb}

This section outlines derivation of the replacement conductivity
$\widehat{\K}^\incl$ and the concentration factor $\widehat{\CH}\sph$
introduced in Eq.~\eqref{eq:single_particle_imperfect}.  Since used in
numerical calculations in~\secref{sec:Numverif}, its two dimensional
format is presented as well. It it shown that both 2D and 3D
concentration factors can be recovered from the solution of a 1D
problem using a simple geometrical argument.

\begin{figure}[ht]
\begin{center}
\begin{tabular}{c@{\hspace{10mm}}c}
\includegraphics*[width=75mm,keepaspectratio]{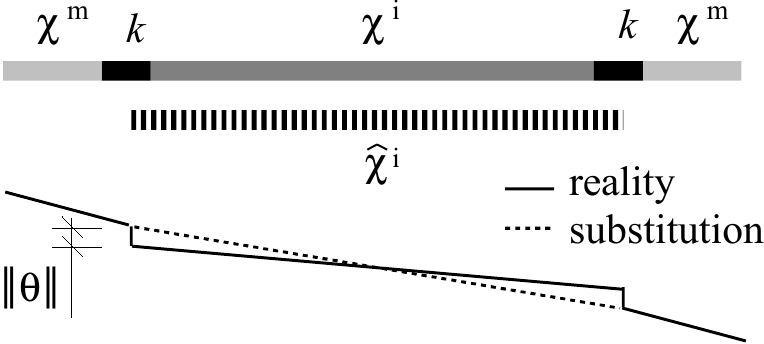}&
\includegraphics*[width=55mm,keepaspectratio]{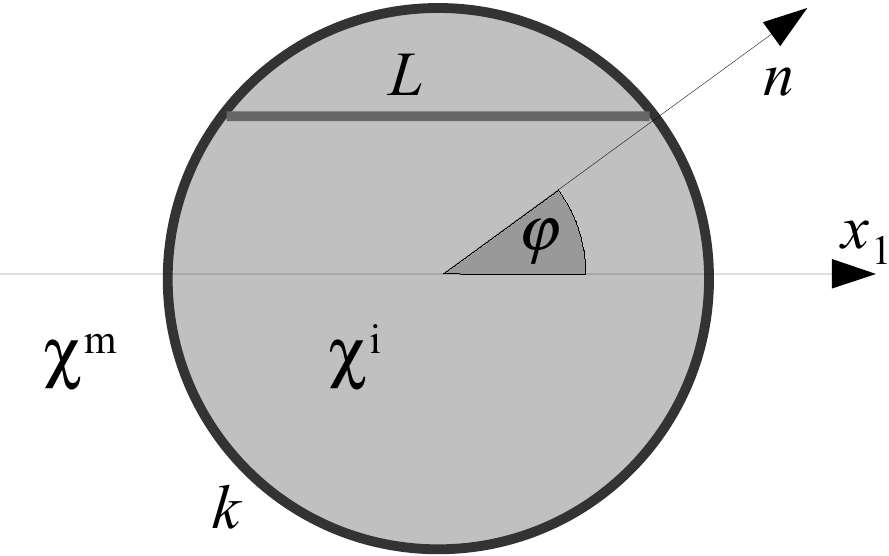}\\
(a)&(b)
\end{tabular}
\end{center}
\caption{Imperfect interface and temperature progress for: a)~1D, b)~2D.}
\label{fig:1D2D_problem}
\end{figure}

To that end, consider one-dimensional heat conduction problem depicted
in~\figref{fig:1D2D_problem}(a). Assuming imperfect thermal contact,
the temperature drop across an infinitely thin interface layer is
given by Eq.~\eqref{eq:temp_jump}. The local temperature gradient for
perfect interface between a solitary inclusion embedded into an
infinite matrix follows from Eq.~\eqref{eq:CF}
\begin{equation}
H^{\rm i} = \mCHi{H} = \frac{\Km}{\Ki}H.
\label{eq:A1D}
\end{equation}
To arrive at similar format of Eq.~\eqref{eq:A1D} for imperfect
contact, we imagine the interface temperature jump being smeared over
the inclusion. Since the heat flux $Q$ associated with the macroscopic
temperature gradient $H$ is constant throughout the composite, we
obtain the total temperature change in the substitute inclusion in the
form
\begin{equation}
\Delta\widehat{\tmp}^\incl = \Delta\tmp^\mathrm{i} + 2\jump{\tmp} = -Q\left(\frac{L}{\Ki} + \frac{2}{k}\right),
\label{eq:deltaT_1D}
\end{equation}
where $L$ stands for the inclusion length. Next, defining the local
temperature gradient in the substitute inclusion as
$H^{\rm i}=\Delta\widehat{\tmp}^\incl/L$ yields the local constitutive
law in terms of the replacement conductivity $\widehat{\K}^\incl$
\begin{equation}
Q = -\widehat{\K}^{\incl}H^{\rm i}\,=\, -\widehat{\K}^\incl \frac{\Delta\widehat{\tmp}^\incl}{L} \,=\, -\frac{\widehat{\K}^\incl}{L} 
\left[-Q\left(\frac{L}{\Ki} + \frac{2}{k}\right)\right],
\end{equation}
so that
\begin{equation}
\widehat{\K}^\incl = \Ki\frac{Lk}{Lk + 2\Ki},
\end{equation}
and in analogy with Eq.~\eqref{eq:A1D} we finally get 
\begin{equation}
H^{\rm i} = \widehat{A}^{\rm i}H = \frac{\Km}{\widehat{\chi}\phs{i}}H.
\label{eq:A1DR}
\end{equation}

The two-dimensional problem of a solitary circular inclusion is
treated similarly. We build up on the assumption that the temperature
gradient in the inclusion is constant and collinear with the
prescribed far field gradient parallel to the local $x_1$-axis,
see~\figref{fig:1D2D_problem}(b). To draw similarity with 1D case we
divide the inclusion into parallel filaments with the length $L =
2a\cos\newtext{\varphi}$. Next, define a unit vector normal to the inclusion
surface $\vek{n}=\left(\cos\newtext{\varphi},\sin\newtext{\varphi} \right)\trn$ and in
analogy to Eq.~\eqref{eq:deltaT_1D}) write the total temperature
change in each filament for the constant heat flux $\vek{q}^\incl =
\left(q^\incl, 0\right)\trn$ as
\begin{equation}
\Delta\widehat{\tmp}^\incl = \Delta\tmp^\mathrm{i} + 2\jump{\tmp} = -q^\incl\left(\frac{d\cos\newtext{\varphi}}{\Ki} + \frac{2\cos\newtext{\varphi}}{k}\right),
\end{equation}
where $d$ is the inclusion diameter. The equivalent conductivity has
to fulfill the condition
\begin{equation}
q^\incl 
= 
-\widehat{\K}^\incl \frac{\Delta\widehat{\tmp}^\incl}{d\cos\newtext{\varphi}} 
= 
-\frac{\widehat{\K}^\incl}{d\cos\newtext{\varphi}}
\left[-q^\incl\left(\frac{2a\cos\newtext{\varphi}}{\Ki} + \frac{2\cos\varphi}{k}\right)\right],
\end{equation}
which yields 
\begin{equation}
\widehat{\K}^\incl = \Ki\frac{ak}{ak + \Ki}.
\label{eq:Keq_3D}
\end{equation}
Consequently, the concentration factor of the substitute inclusion
attains the form
\begin{equation}
\widehat{A}^{\rm i} = \frac{2\Km}{\Km+\widehat{\K}^\incl}.
\end{equation}

The analysis of a spherical inclusion follows identical steps. The
replacement thermal conductivity for constant heat flux $\vek{q}^\incl
= \left(q^\incl, 0, 0\right)\trn$ thus receives the same form as
in~Eq.~\eqref{eq:Keq_3D} rendering the searched concentration factor
as, recall Eqs.~\eqref{eq:general_imperfect_system},
\begin{equation}
\widehat{A}^{\rm i} = \frac{3\Km}{2\Km+\widehat{\K}^\incl}.
\end{equation}

\end{document}

%% file: format.tex
\usepackage{stmaryrd}

\newcommand{\figref}[1]{\figurename~\ref{#1}}
\newcommand{\secref}[1]{Section~\ref{#1}}
\newcommand{\tabref}[1]{\tablename~\ref{#1}}

\newcommand\de{\,{\mathrm d}} 

\newcommand{\vek}[1]{\boldsymbol{#1}}
\newcommand{\trn}{{\sf ^T}}
\newcommand{\loc}{_{\ell}}

\newcommand{\x}{\vek{x}}
\newcommand{\MH}{\vek{H}}
\newcommand{\mH}{\vek{h}}
\newcommand{\MQ}{\vek{Q}}
\newcommand{\mQ}{\vek{q}}
\newcommand{\mK}{\vek{\chi}}
\newcommand{\MK}{\vek{\chi}^\mathrm{H}}

\newcommand{\incl}{{\mathrm{i}}}
\newcommand{\mtrx}{{\mathrm{m}}}

\newcommand{\CHi}{\CH^\incl}
\newcommand{\sph}{_{\mathrm{sph}}}
\newcommand{\cir}{_{\mathrm{cir}}}

\newcommand{\Oi}{\Omega^\incl}
\newcommand{\Om}{\Omega^\mtrx}

\newcommand{\mKi}{\mK^\incl}
\newcommand{\mKm}{\mK^\mtrx}

\newcommand{\tmp}{\theta}
\newcommand{\n}{\vek{n}}

\newcommand{\mCH}{\vek{A}}

\newcommand{\mCHi}{\mCH^\incl}
\newcommand{\I}{\vek{I}}
\newcommand{\Et}{\vek{S}}
\newcommand{\Km}{\chi^\mtrx}
\newcommand{\Ki}{\chi^\incl}

\newcommand{\phs}[1]{^{(#1)}}
\newcommand{\MHm}{\MH^\mtrx}
\renewcommand{\O}{\Omega}
\newcommand{\meas}[1]{\left|#1\right|}
\newcommand{\vfrac}{c}

\newcommand{\oavg}[1]{\langle\!\langle #1 \rangle\!\rangle}

\newcommand{\K}{\chi}
\newcommand{\KH}{\chi^\mathrm{H}}
\newcommand{\CH}{A}
\newcommand{\jump}[1]{\llbracket #1\rrbracket}

\newcommand{\rd}{a}
\newcommand{\mCHr}[1]{\vek{T}\phs{#1}}

\newcommand{\pCH}[1]{T\phs{#1}}

\newcommand{\iCH}[1]{\widehat{T}\phs{#1}}
\renewcommand{\d}{\, \mathrm{d}}
\newcommand{\savg}[1]{\left\{ #1 \right\}}

\newcommand{\ar}{\beta} 